\documentclass[sigconf]{acmart}

\usepackage{color, xcolor, float, lscape, enumerate, graphicx, url, tabularx, multirow, xspace}
\usepackage{hyperref}
\usepackage{blindtext}
\usepackage{booktabs}
\usepackage{etoolbox}
\usepackage{multirow}
\usepackage{amsfonts}
\usepackage[linesnumbered,ruled]{algorithm2e}
\usepackage{wrapfig}
\usepackage{enumitem}

\usepackage{pdfpages}
\usepackage{comment}
\usepackage{amsmath,amssymb,amsfonts}
\usepackage{graphicx}
\usepackage{times}
\usepackage[ruled,linesnumbered]{algorithm2e}
\usepackage{multicol}
\usepackage{multirow}
\usepackage{xspace}
\usepackage{xcolor}
\usepackage{balance}
\usepackage{subcaption}

\hypersetup{linktocpage}
\usepackage{array}
\usepackage{tikz}
\usepackage{rotating}
\usetikzlibrary{arrows}
\usepackage{verbatim}

\usepackage{balance}

\newcommand{\BfPara}[1]{{\noindent\bf#1.}\xspace}

\newcommand{\etal}{{\em et al.}\xspace}
\newcommand{\eg}{{\em e.g.,}\xspace}
\newcommand{\ie}{{\em i.e.,}\xspace}
\newcommand{\etc}{{\em etc.}\xspace}
\newcommand{\ssmc}{DL-SSMC}
\newcommand{\fhmc}{DL-FHMC}

\definecolor{dkgreen}{rgb}{0,0.6,0}
\definecolor{gray}{rgb}{0.5,0.5,0.5}
\definecolor{mauve}{rgb}{0.58,0,0.82}

\newcommand*\cib[1]{\tikz[baseline=(char.base)]{
                            \node[shape=circle,color=black, fill=black,text=white,draw,inner sep=0.3pt] (char) {#1};}}

\usepackage{listings}

\setcopyright{none}
\renewcommand\footnotetextcopyrightpermission[1]{}
\begin{document}

\title{A Deep Learning-based Fine-grained Hierarchical Learning Approach for Robust Malware Classification}

\author{Ahmed Abusnaina}
\affiliation{%
  \institution{University of Central Florida}}
\email{ahmed.abusnaina@knights.ucf.edu}

\author{Mohammed Abuhamad}
\affiliation{%
  \institution{University of Central Florida}}
\email{abuhamad@knights.ucf.edu}

\author{Hisham Alasmary}
\affiliation{%
  \institution{University of Central Florida\\King Khalid University}}
\email{hisham@knights.ucf.edu}

\author{Afsah Anwar}
\affiliation{%
  \institution{University of Central Florida}}
\email{afsahanwar@knights.ucf.edu}

\author{Rhongho Jang}
\affiliation{%
  \institution{University of Central Florida}}
\email{r.h.jang@knights.ucf.edu}

\author{Saeed Salem}
\affiliation{%
  \institution{North Dakota State University}}
\email{saeed.salem@ndsu.edu}

\author{DaeHun Nyang}
\affiliation{%
  \institution{Ewha Womans University}}
\email{nyang@ewha.ac.kr}

\author{David Mohaisen}
\affiliation{%
  \institution{University of Central Florida}}
\email{mohaisen@ucf.edu}

\renewcommand{\shortauthors}{Abusnaina, et al.}

\begin{abstract}
The wide acceptance of Internet of Things (IoT) for both household and industrial applications is accompanied by several security concerns. A major security concern is their probable abuse by adversaries towards their malicious intent. 
Understanding and analyzing IoT malicious behaviors is crucial, especially with their rapid growth and adoption in wide-range of applications. Among the variety of employed techniques, static and dynamic analyses are the most common approaches to detect and classify malware.
Given the limited scalability of dynamic analysis, static analysis, such as the use of Control Flow Graph (CFG)-based features, is widely used by machine learning algorithms for malware analysis and detection. 
However, recent studies have shown that machine learning-based approaches are susceptible to adversarial attacks by adding junk codes to the binaries, for example, with an intention to fool those machine learning or deep learning-based detection systems. 
Realizing the importance of addressing this challenge, this study proposes a malware detection system that is robust to adversarial attacks. 
To do so, examine the performance of the state-of-the-art methods against adversarial IoT software crafted using the graph embedding and augmentation techniques. In particular, we study the robustness of such methods against two black-box adversarial methods, GEA and SGEA, to generate Adversarial Examples (AEs) with reduced overhead, and keeping their practicality intact.
Our comprehensive experimentation with GEA-based AEs show the relation between misclassification and the graph size of the injected sample. Upon optimization and with small perturbation, by use of SGEA, all the IoT malware samples are misclassified as benign. 
This highlights the vulnerability of current detection systems under adversarial settings. 
With the landscape of possible adversarial attacks, we then propose \fhmc{}, a fine-grained hierarchical learning approach for malware detection and classification, that is robust to AEs with a capability to detect 88.52\% of the malicious AEs.
\end{abstract}


\maketitle
\section{Introduction}\label{sec:introduction}
The Internet of Things (IoT) has shown its fast growth in the last decade. The communication media of IoT devices is not limited only in the form of home networking, but also the cellular, which significantly accelerated their connectivity and accessibility. According to Ericsson~\cite{Erisson}, 3.5 billion IoT devices are expected to communicate using cellular in 2023. On the one side, taking advantage from the a large number of interconnected devices, many applications can be adopted on a large scale. Moreover, due to the high-speed and low-latency connection of these devices, time critical applications will become feasible in the real world. On the down side, however, the high accessibility also provides a convenience to adversaries. Due to the constrained resource of IoT devices, the protection function is usually inefficient. 
Moreover, due to the large population, low physical accessibility, and the unrestricted use of policies in many scenarios, IoT devices can be easily compromised and abused by adversaries to launch a variety of attacks, such as Distributed Denial of Service (DDoS) attacks launched by Mirai botnet~\cite{AntonakakisABB17}. Such adversarial scenarios are expected to grow for many years to come, posing a critical security threats for the IoT ecosystem.

There has been a large body of research work on the problem of malware analysis using both static and dynamic approaches~\cite{MohaisenAM15,Gerber17,JiaCWRFMP17}, and a few attempts on analyzing IoT malware in particular. Recently, machine learning algorithms, specifically deep learning techniques, are actively utilized in security and privacy applications~\cite{Cui18,JuarezAADG14,abuhamad2018large, abuhamad2019code, abuhamad2020autosen}, particularly, for detecting and classifying malicious software from benign ones. However, it should be noted that the research work on IoT malware analysis has been very limited not only in the size of the analyzed samples, but also the utilized approaches~\cite{azmoodehDCK18,DemontisMBMARCG19,SibyMT17,SaracinoSDM18}.
Among the static analysis-based approaches, one of the prominent approach is to use an abstract graph structures for IoT malware analysis and detection, such as the control flow graph (CFG)~\cite{AlasmaryKAPCAANM19,AlasmaryAJAANM20}. Previously, it has been shown that the software graph-based analysis can be incorporated with machine learning methods to introduce more powerful analysis tools~\cite{yin2007panorama,antonakakis2012throw}.
For the IoT malware detection, CFGs allow defenders to extract plentiful feature representations that can be used to distinguish those malware from benign, owing to their various properties, such as the degree distribution, centrality measures, radius,~\etc.~\cite{AlasmaryKAPCAANM19}. 
Those properties can be represented as a feature vector that can be used to enable machine learning algorithms to accurately detect and classify IoT malware samples. Proposed by Alasmary~\etal, one such application is exploring IoT malware using both graph analysis and machine/deep learning~\cite{AlasmaryKAPCAANM19}. Their model not only can learn the representative characteristics of the graph, but also can be utilized to build an automatic detection system for predicting the label of the unseen software.

Unfortunately, machine learning-based IoT malware detection methods are prone to adversarial manipulation~\cite{PapernotMGJCS17}. 
The rise of adversarial machine learning has highlighted the fragile nature of those algorithms to perturbation and data poisoning attacks that lead to misclassification.
For example, an adversary can introduce a small modification to the input sample to make the classifier misidentify the malware as a benign (\ie the adversary introduce an adversarial example). Such modification are usually crafted using small perturbation to make the 
AE undetectable and very difficult to distinguish for the original sample.

To date, there has been a large body of work exploring the generation of AEs in general image-based classification problems~\cite{GoodfellowSS15,Moosavi-Dezfooli16} as well as in the context of malware classification~\cite{PapernotMJFCS16,GrossePMBM17,AbusnainaKAPAM}.
Also, there is an active trend in the research area towards investigating adversarial machine learning to overcome these threat. 
However, there is very limited research that aims to understand the impact of adversarial learning on deep learning-based IoT malware detection system and practical implications, especially for those approaches that utilize CFG features.
We stress the importance of addressing the threat posed by the machine learning vulnerability to AEs, particularly in security-sensitive applications. 
We undertake this challenge by \cib{1} showing the high potential of successful detection/classification of IoT malware using deep learning methods; \cib{2} assessing the robustness of such methods to AEs generated by different state-of-the-art CFG-based AEs generation techniques; \cib{3} introducing a fine-grained hierarchical approach to tackle adversarial attacks by leveraging patterns extracted from the basic and elementary structure of the tested software.

To this end, we introduce \ssmc{}, \textbf{D}eep \textbf{L}earning-based \textbf{S}ingle \textbf{S}hot \textbf{M}alware \textbf{C}lassification approach, for accurate IoT malware detection and classification. 
Then, we examine the approach against AEs generated by GEA~\cite{AbusnainaKAPAM} and SGEA~\cite{AbusnainaAASNM19}. 
The GEA and SGEA are graph-based AEs generation approaches that are recently proposed to launch targeted and untargeted misclassification attacks.
To cope with adversarial settings and to minimize their effects, we propose \fhmc{}, \textbf{F}ine-grained \textbf{H}ierarchical Learning for \textbf{M}alware \textbf{C}lassification, for detecting and classifying malware samples by operating on a fine-grained level of structures and patterns extracted from the tested software sample.
Our experiments show the effectiveness of the proposed approach in detecting malware samples as well as high-degree of robustness against variety of adversarial attacks.

\BfPara{Summary of contributions} Our contributions are as follows: First, we introduce \ssmc{} for IoT malware detection and classification.
Second, we examine the robustness of the CFG-based deep learning IoT malware detection system using two different approaches, including GEA and SGEA. 
The GEA approach generates adversarial IoT software, through embedding representative target sample to the original software, while maintaining the practicality and functionality of the targeted sample. 
SGEA is an optimization of GEA to generate AEs with reduced injection size. 
Through comprehensive experiments, we show the effectiveness of GEA and SGEA in producing successful AEs that can fool the machine learning-based malware detection system. 
The results show that GEA and SGEA approaches achieve a malware-to-benign misclassification rate of up to 100\%.
To tackle such vulnerability, we propose \fhmc{} that gains insights by 
investigating the malicious subgraphs (patterns) within the IoT malware families.
Using a fine-grained hierarchical learning approach, \fhmc{} utilizes subgraph mining and pattern recognition to detect suspicious and malicious behaviors within the tested samples, mitigating the effects of AEs and detecting 88.52\% of malicious AEs.

\BfPara{Organization}\label{sec:organization}
This work is organized as follows: In~\textsection\ref{sec:preliminaries}, a brief background is provided. The presented practical approach for generating practical adversarial IoT software is described in~\textsection\ref{sec:methodology}. In~\textsection\ref{sec:No_defense}, we evaluate the proposed adversarial methods on traditional deep learning techniques. Then, we propose a fine-grained hierarchical learning technique for suspicious behavior detection in~\textsection\ref{sec:sbd}. In~\textsection\ref{sec:discussion}, we discuss the proposed adversarial methods and suspicious behavior detection approach. 
Related work has been discussed in~\textsection\ref{sec:Related_Work}. Finally, we conclude our work in~\textsection\ref{sec:conclusion}. 

\section{Preliminaries}\label{sec:preliminaries}
We incorporate adversarial learning techniques into CFG-based deep learning IoT malware detection systems in an attempt to understand the robustness of such models against adversarial learning attacks as a result of AEs. 
We provide a required preliminary knowledge for understanding those techniques and approaches required for malware analysis to extract graph structures and to automate their labeling using machine learning. In particular, we provide general knowledge about the malware analysis approaches in~\textsection\ref{sec:MA}. The CFG-based analysis for IoT malware detection is described in~\textsection\ref{sec:GA}. Finally, we describe background knowledge about the concept of adversarial machine learning and its effects on machine learning models in~\textsection\ref{sec:AADL}.

\subsection{Malware Analysis}\label{sec:MA}
Malware analysis is widely used to understand the functionality and behavior of malware. It helps us to understand the capabilities and the intent of the malware and malware authors. The results of the analyses are often used to build detectors and design defenses to protect against future malware campaigns. There are two approaches utilized for analyzing malicious software: (i) static and (ii) dynamic analyses. Static analysis approaches analyze malware binaries without the need for execution. Given the malicious nature of the malware, static analysis is utilized as a precursor to the dynamic analysis. The malware binary can then be executed in a sandboxed environment with a much-reduced focus to observe the patterns, like the behavioral artifacts in which is called the dynamic analysis. 

\BfPara{Static Analysis} 
Static analysis approaches employ various techniques to extract indicators to determine whether the software is malicious or benign~\cite{zhangR07}. The various analysis points, such as instructions, basic blocks, functions \etc hint at the possible execution pattern of the software. 
For example, the traces of using user name and password list, along with shell-based login attempts, implies possible usage of dictionary attack being launched by the software. These inferential results are drawn from static analysis to enable the analysts to emphasize and predict specific patterns. 
Additionally, traces can also be used by analysts to address issues during dynamic analysis, \eg virtual machine obfuscation, ptrace obfuscation \etc.
Traces and components of software are often extracted using a reverse engineering process of the software binaries that allow the understanding of its composition, architecture, and design.
Analysts perform reverse engineering of software binaries using several available off-the-shelf tools.
The reverse engineering process also generates artifacts to be subject of analysis including high-level representation of the binaries such as the CFGs and Data Flow Graphs (DFGs).
The CFG  of a program is the graphical representation of the flow of control during the execution of that program.
While the DFG represents the system events to understand the possible execution of the system behaviors. It explains the flow of the data that passes from one node to another. 
Although static analysis is quite powerful and popular, it sometimes falls short of achieving its end goals of providing in-depth insights to the software due to multiple evasion techniques. 
For example, malware developers use evasion techniques to the analysis of their produced malware. The evasion techniques include packing (UPX~\cite{upx}), obfuscation (function-, string- obfuscation), \etc.

\BfPara{Dynamic Analysis}
Unlike static analysis, dynamic analysis executes the application in a simulated and monitored environment to observe its behavior and understand its functionality~\cite{willemsHF07}. This approach aims to capture different behavioral patterns of software. For example, dynamic analysis helps unraveling the program's network patterns, such as communication with the Command and Control (C\&C) server.  Since the malicious nature of software can affect the status of the machine it is executed on, the following observations are made: 1) comparing the system state before and after the execution of the application, or 2) monitoring the application's actions during the execution. 
Similar to the static analysis, dynamic analysis can be evaded by software developers by adopting means that prevent their software from getting dynamically-analyzed. For example, malware developers often employ conditions that crashes the software once encountering virtual machines, debugging tools or network monitoring tools.

\subsection{Graph Analysis}\label{sec:GA}
\BfPara{Graph Analysis}
The CFG is a graph representation of the program which shows all paths that can be reached during the execution as in \autoref{fig:GC_Xorg}. In a CFG, the set of nodes means the basic blocks where each block is a straight-line instruction without any {\em jump} or {\em jump target}, while the set of directed edges corresponds to the {\em jump} which traverses from the block to the other block at the branch ({\em if}), loop ({\em while, for}), and the end of the function ({\em return}). Once the first instruction of the basic block is executed, the rest of the instructions in the same block are necessarily executed unless terminated by external interference. In general, CFG is used for the structural analysis of the program. For example, from the perspective of optimization, the CFG is used to analyze the reachability of each block. By constructing the CFG and evaluating the reachability, the flaws of the program (infinite loop or unreachable codes) can be found and addressed. 

\BfPara{CFG-based Analysis}
In graph theory, there are various concepts that express the characteristics of a graph. Given $G=(V,E)$, for example, the number of vertices ($|V|$) means the order of $G$, while the number of edges ($|E|$) corresponds to the size of $G$. The density of the graph can be defined as $D=|E|/(|V|*(|V|-1))$ for directed simple graph, which means the ratio of the number of edges in $G$ to the maximal number of edges in the complete graph. The centrality is measured for each node $v \in V$, which shows how important a specific node is. In detail, there are several different kinds of centrality, such as closeness centrality, betweenness centrality, Eigenvector centrality, \etc. 

These indicators (and further concepts not described above) can be considered the features of the graph $G$. Moreover, the combination of those metrics can be a more deterministic characteristic of the graph. Considering that a CFG is a kind of graph, it is true that each binary has not only its unique graph representation but also the associated values, such as the order, size, and density of CFG, and centrality for each vertex in CFG.  On the other hand, the graph-based analysis can provide the possibility for identifying the malware. Because it is highly likely that the binaries in the same ``family'' share the structural similarity (even if there is a little difference), the CFG-based features can be combined with the state-of-the-art machine learning technique to determine whether a given binary is malicious.

\subsection{Threat Model}\label{sec:AADL}
The rapid reliance on machine learning methods in various applications have raised several security and privacy concerns, especially in security-sensitive applications.
It has become crucial to understand and assess the robustness of machine learning techniques to several adversarial settings.
These adversarial settings includes AEs, where 
an adversary intends to fool or misguide the classification model with malicious inputs that are generated by applying minimal perturbation to the original sample~\cite{PapernotMJFCS16}. 
These inputs misclassify the samples of the model from benign to malware and vice versa and even misclassify the malware classes to another class. 
Such adversarial attacks can be launched under different adversarial capabilities that allow for either black-box and white-box attack. In a white-box attack, the adversary has full knowledge of the inner networking paradigm of the model, while in a black-box attack, the adversary has only access to the model via an oracle and observe only the output of the model.

The literature on AEs and their effects shows numerous studies conducted on images, where the perturbation is applied to image pixels~\cite{PapernotMJFCS16,PapernotMGJCS17,wangYVZZ18}.
Unlike image AEs, the generated AEs from the IoT software must preserve the original sample's functionality and practicality in order to function properly. 
In this study, we generate AEs from the IoT software based on code-level manipulation using GEA~\cite{AbusnainaKAPAM} and SGEA~\cite{AbusnainaAASNM19}. 
Adversarial machine learning can be derived from two perspectives: targeted and non-targeted attacks.
\BfPara{Targeted attacks} 
The focus of this attack is to generate AE $x'$ that forces the classifier $f$ to misclassify into a specific target class $t$. For instance, the adversary generates a set of malicious IoT software samples, which are classified as benign. That is:
   $ x': [f\left ( x' \right) = t] \wedge [ \Delta \left ( x , x' \right) \leq \epsilon ]$,
where $f(.)$  represents the classifier’s output, $\Delta \left ( x , x' \right)$ denotes the difference between $x$ and $x'$, whereas $\epsilon$ is the distortion threshold.

\BfPara{Non-targeted attacks} 
The focus of non-targeted attack is to generate an AE that forces the classifier $f$ to misclassify to any class other than the original class $f(x)$, where $x$ is the original input. That is:
 $x': [f\left ( x' \right) \neq f\left ( x \right)] \wedge [\Delta \left ( x , x' \right) \leq \epsilon]$, 
where $f(.)$ shows the classifier's output, $\Delta \left ( x , x' \right)$ represents the difference between $x$ and $x'$, and  $\epsilon$ is the distortion threshold.

\section{Malware and Adversarial Examples}\label{sec:methodology}
This work aims to investigate the robustness of current practices of machine learning in analyzing IoT malware for detection and classification, especially against sophisticated AEs.
To generate AE, an adversary applies a perturbation $\epsilon$ to the model's input $x$ to fool the classifier by generating a new input feature space $x'=x+\epsilon$. Thus, the generated output $f$ of the model $f(x) \neq f(x')$, aiming to misclassify the output of the targeted model.
For a practical and functional adversarial behavior with the ability of being hiding from detection systems, the perturbation $\epsilon$ is minimized to be $\delta=\epsilon_{min}$, where $f(x)\neq f(x+\delta)$.
For IoT malware detection system, the AE process aims to generate AEs that preserve the functionality and practicality of the original IoT software.
Recent work \cite{AbusnainaKAPAM} proposed a Graph Embedding and Augmentation (GEA) approach that ensures the practicality of the AEs by applying $\epsilon$ at the code-level of the IoT samples. 
To reduce the perturbation overhead of GEA for efficient adversarial settings, another approach for generating AEs, Sub-GEA (SGEA)~\cite{AbusnainaAASNM19}, was proposed. 
This section describes the dataset used in this work and the two approaches for generating AEs, namely: GEA and SGEA.

\begin{table}[t]
    \centering
    \caption{Distribution of IoT samples across the classes. We split the dataset into 80\% training and 20\% testing, with an overall 8,670 IoT samples (5,670 IoT malware and 3,000 benign).}
    \label{DSDist}
    \begin{tabular}{l|l|c|c|c|c}
        \toprule
        \multicolumn{2}{c|}{\multirow{2}{*}{Class}} & \multicolumn{3}{c|}{\# of Samples} & \multirow{ 2}{*}{\% of Samples}\\
        \multicolumn{2}{c|}{} & \# Train & \# Test & \# Total \\
        \midrule
        \multicolumn{2}{c|}{Benign} & 2,400 & 600 &  3,000 & 34.60\%  \\ 
        \midrule
        \multirow{3}{*}{Malicious} & Gafgyt & 2,400 & 600 &  3,000 & 34.60\%  \\ 
         & Mirai & 1,927 & 481 &  2,408 & 27.68\%  \\ 
         & Tsunami & 210 & 52 &  262 & 3.02\%  \\ 
        \midrule
        \multicolumn{2}{c|}{Overall} & 6,937 & 1,733 & 8,670 & 100\% \\
        \bottomrule
    \end{tabular}
\end{table}

\begin{table}[t]
    \centering
    \caption{The distribution of extracted features. 23 algorithmic features are extracted from the CFGs. These features are categorized into seven groups, including number of nodes and edges, density, shortest path, and centralities. When possible, the minimum, maximum, median, mean, and standard deviation values are extracted, as in the shortest path group.}
    \label{DSFeatures}
    \begin{tabular}{l|c}
        \toprule
        Feature category & \# of features\\
        \midrule
        
        Betweenness centrality &  5  \\ 
        Closeness centrality &  5  \\ 
        Degree centrality &  5  \\ 
        Shortest path &  5  \\ 
        Density &  1  \\ 
        \# of Edges &  1  \\ 
        \# of Nodes &  1  \\ 
        
        \midrule
        Total & 23 \\
        \bottomrule
    \end{tabular}
\end{table}
\subsection{Dataset} \label{sec:Dataset}
For the purpose of this study, we gathered a dataset of IoT malware samples. 
We collected binaries of two categories, IoT malicious and benign samples. The malicious samples are collected from CyberIOCs~\cite{cyberiocs19} in the period of January 2018 to late February of 2019, with a total of 5,670 samples that belong to three malware families. Additionally, we assembled a dataset of 3,000 benign IoT samples compiled from the source files on GitHub~\cite{github19}. 

\BfPara{Ground Truth Class} The benign and malicious samples in our dataset were validated using the {\em VirusTotal}~\cite{VirusTotal}. We uploaded the samples on VirusTotal and gathered the scan results corresponding to each sample. We then used AVClass~\cite{SebastianRKC16} to classify the malicious samples into their families. We summarize the dataset in table \ref{DSDist}.

\BfPara{Dataset Creation} Samples of the IoT benign category (3,000 sample) and IoT malware category (5,670 samples) were reverse-engineered using {\em Radare2}~\cite{radare2}, a reverse engineering framework that provides various analysis capabilities, for obtaining the samples' corresponding CFGs. 
Using the samples' CFGs extracted by \textit{Radare2}, we extracted different graph-theoretic features to represent the software sample for the machine learning algorithm. We extracted 23 different algorithmic features categorized into seven groups. ~\autoref{DSFeatures} shows the feature category and the number of features in each category. The five features extracted from each feature category represent minimum, maximum, median, mean, and standard deviation values for the observed parameters.

\begin{table}
  \noindent\begin{minipage}{.45\columnwidth}
\begin{lstlisting}[caption={C script of an example of original sample},captionpos=b,label={lst:XORG_methedology},language=C]
#include <stdio.h> 
void main(){ 
    int a = 0; 
    do
    { 
    	a++; 
    }while(a < 10); 
} 

\end{lstlisting}

      \end{minipage}
      \begin{minipage}{.45\columnwidth}
       
\begin{lstlisting}[caption={C script of an example of targeted sample},captionpos=b,label={lst:XSEL_methedology},language=C]
#include <stdio.h> 
void main(){ 
    int x = 0; 
    int s = 0;
    if(x!=0){
        s++;
    }
} 
\end{lstlisting}

      \end{minipage}
      \begin{minipage}{.96\columnwidth}
       
\begin{lstlisting}[caption={C script of combining original and selected samples. Note that a condition variable is used to enable the desired functionality. Additionally, the script of the original and selected samples are preserved within the generated sample.},captionpos=b,label={lst:XCOMP_methedology},language=C]
#include <stdio.h> 
void main(){ 
    /*set a condition variable*/
    int cond=1;
    if(cond==1){
        /*script of original sample*/
        /*this section will be executed*/
        int a = 0;
        do{
            a++;
        }while(a<10);
    }
    else{
        /*script of target sample*/
        /*this section will not be executed*/
        int x = 0;
        int s = 0;
        if(x!=0){
            s++;
        }
    }
}
\end{lstlisting}

      \end{minipage}
    \end{table}

\subsection{Graph Embedding and Augmentation (GEA)} \label{sec:attack_GEA}
Recent work~\cite{AbusnainaKAPAM} introduces GEA approach that generates realistic AEs, where they maintain the functionality and practicality of the original IoT sample and achieving a high misclassification rate. The main goal was to combine an original CFG with a selected targeted CFG. The authors show that such a combination leads to a high misclassification rate. In the following, we briefly describe the GEA approach using an example.

\begin{figure*}
\centering
\begin{minipage}[t]{0.40\textwidth}
\includegraphics[width=0.99\textwidth]{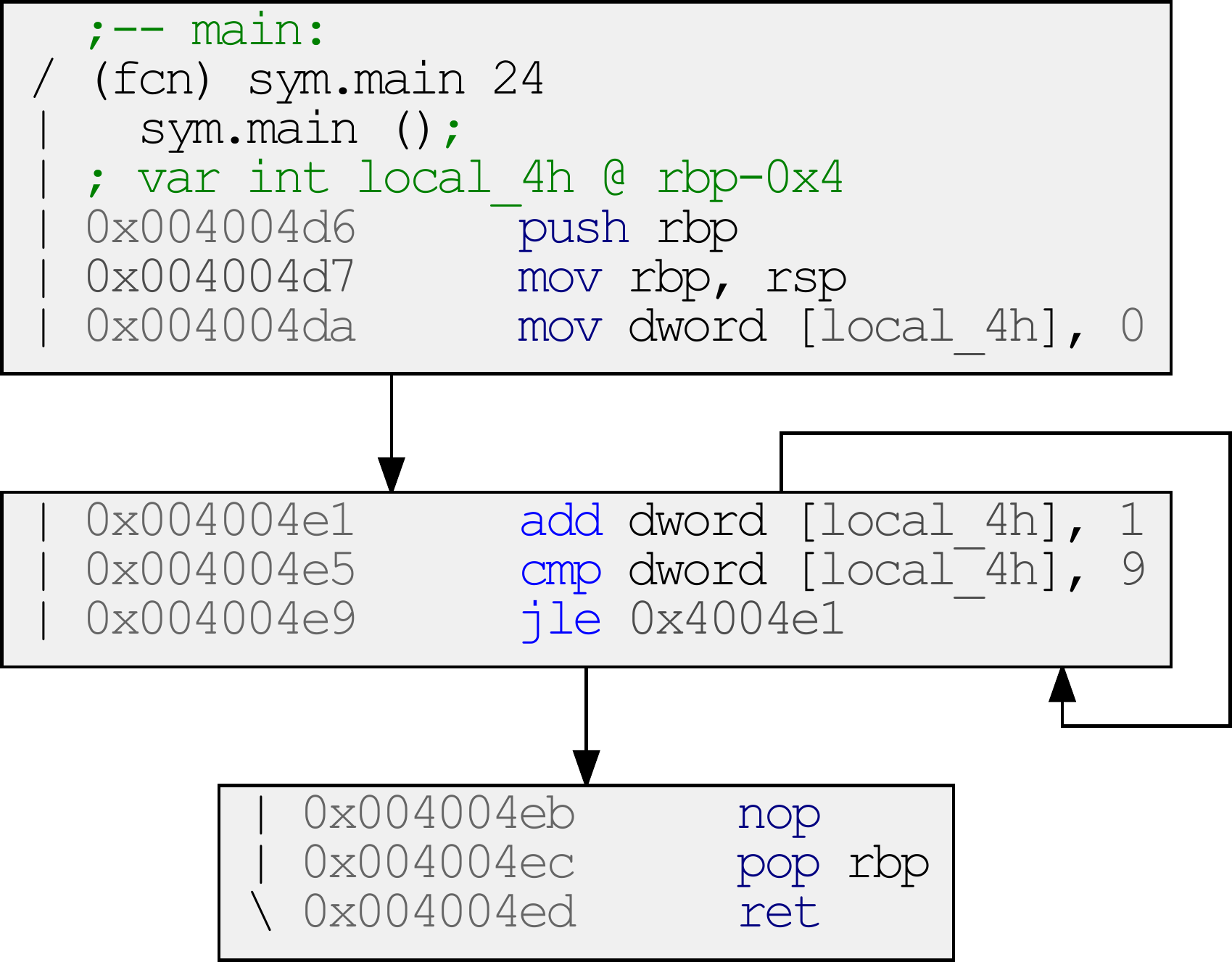}
\caption{The generated CFG for the original sample and used for extracting graph-based features (graph size, centralities, etc.) for graph/program classification and malware detection.}

\label{fig:GC_Xorg}
\end{minipage}
\hfill
\begin{minipage}[t]{0.47\textwidth}
\includegraphics[width=0.99\textwidth]{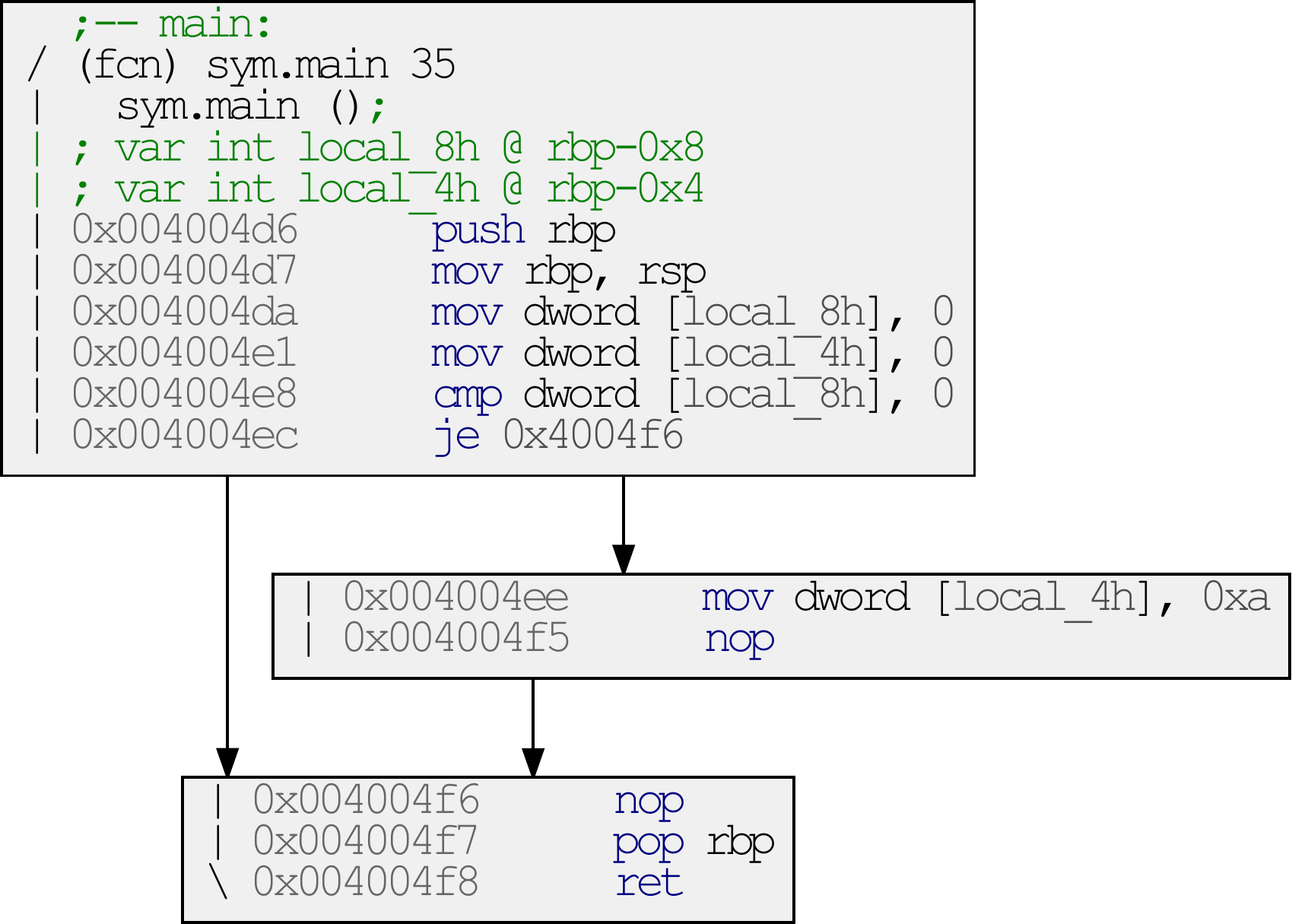}
\caption{The CFG for the selected target sample generated and used for extracting graph-based features (graph size, centralities, etc.) for graph/program classification and malware detection.}
\label{fig:GC_Xsel}
\end{minipage}
\end{figure*}

\begin{figure}
\centering
\includegraphics[width=0.48\textwidth]{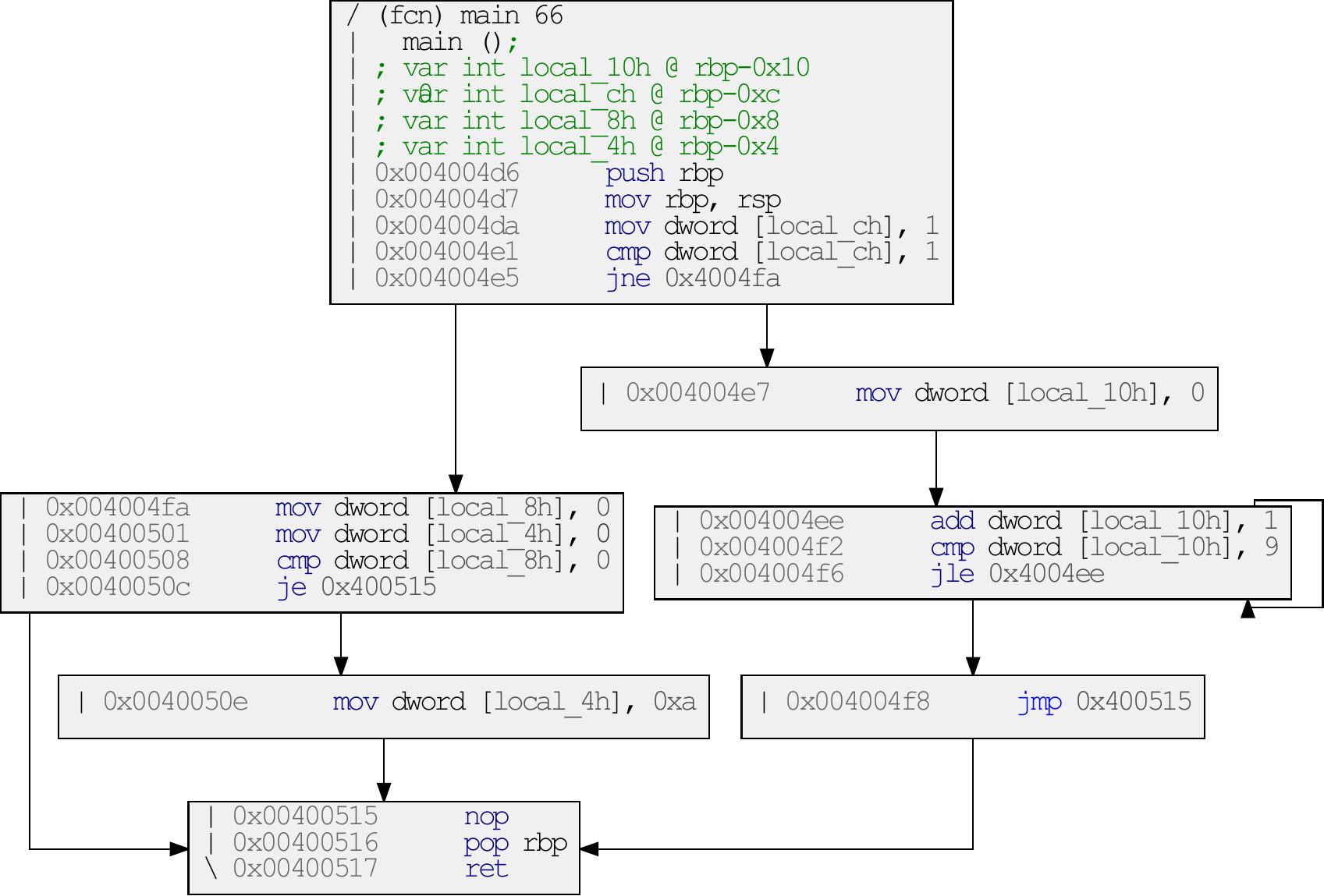}
\caption{The generated adversarial graph using GEA approach. Note that this graph is obtained logically by embedding the graph in Fig.~\ref{fig:GC_Xsel} into the graph in Fig.~\ref{fig:GC_Xorg}, although indirectly done by injecting the code listings in Listings 1, 2, and 3.}
\label{fig:GC_Xcomp}
\end{figure}

\BfPara{Practical Implementation} \label{sec:GC_BE} 
Assume $x_{org}$ is an original sample and $x_{sel}$ is a selected target sample, GEA aims to combine the original sample ($x_{org}$) with a selected target sample ($x_{sel}$) while preserving the functionality and practicality of the original sample ($x_{org}$). 
For example, \autoref{lst:XORG_methedology} shows the original sample script ($x_{org}$), and \autoref{lst:XSEL_methedology} shows the selected target sample script ($x_{sel}$). The GEA process combines the two scripts while ensuring that ($x_{sel}$) does not affect the process and functionality of ($x_{org}$). The generated AE in~\autoref{lst:XCOMP_methedology} shows the script after the combination process. Note that the condition is set to execute only the functionality related to ($x_{org}$) and preventing the processes of ($x_{sel}$) from being executed. Prior to generating the CFG for these algorithms, we compile the code using the GNU Compiler Collection (GCC) command. Afterward, Radare2~\cite{radare2} is used to extract the CFG from the binaries. \autoref{fig:GC_Xorg} and \autoref{fig:GC_Xsel} show the generated CFGs for both ($x_{org}$) and ($x_{sel}$), respectively. As shown in~\autoref{fig:GC_Xcomp}, the combined CFG consists of the two scripts sharing the same entry and exit nodes.
Therefore, the GEA approach adds modifications to the CFG for generating the AE. Given the nature of the extracted features, the applied changes on the CFG is reflected upon the features, regardless of the effects on the functionality and executability of the original sample. 
Following the adopted approach in \cite{AbusnainaKAPAM}, we select six different-sized graphs from benign and malicious samples as $x_{sel}$. The selected graphs vary in size by minimum, median, or max graph size, where the size is the number of nodes in the graph. To generate AEs, we selected a graph and connected it with all samples from the opposite class using shared entry and exit nodes.

\begin{figure}
\centering
\includegraphics[width=0.12\textwidth]{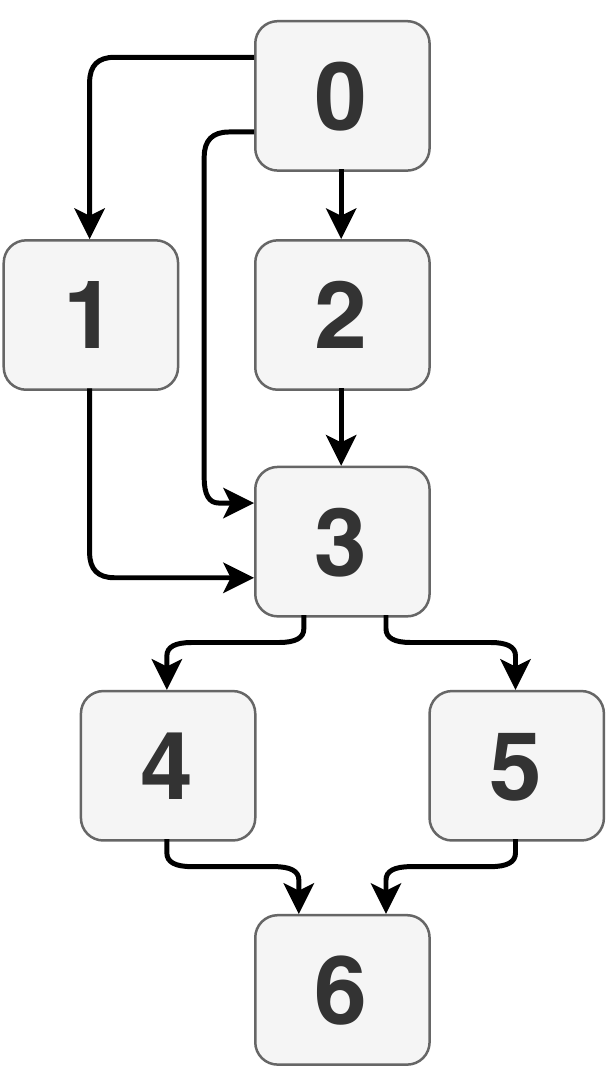}
\caption{Sample of extracted discriminative subgraph from Gafgyt malicious family. Here, the graph size is 7, and the labels are arbitrary. Ideally, connecting this subgraph to a sample should lead the model to misclassify the sample into Gafgyt.}
\label{fig:SGEA_Pattern}
\end{figure}

\begin{figure*}[t]
\centering
\includegraphics[width=0.75\textwidth]{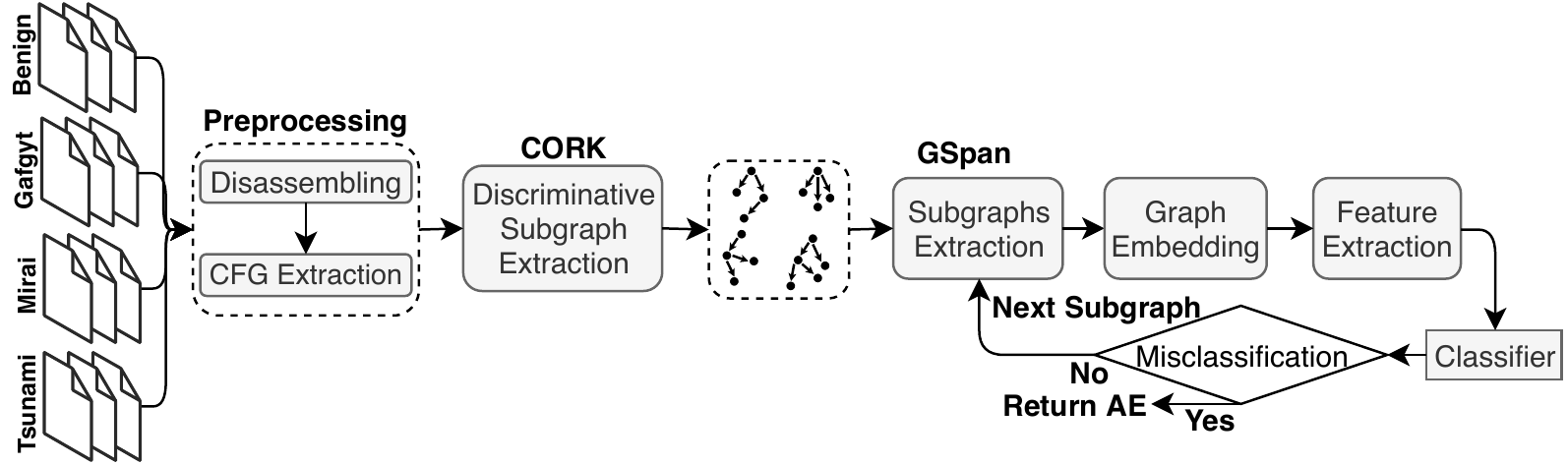}
\caption{SGEA pattern extraction and AE generation process. SGEA uses CORK to extract discriminative subgraphs from each class. Then, the extracted subgraphs are embedded to generate the AEs. The process ends when the generated AE misclassifies the model.}
\label{fig:PatternExtraction}
\end{figure*}

\subsection{Sub-GEA (SGEA)}\label{sec:attack_SGEA}
While GEA combines an original CFG to a selected CFG of the IoT samples to fool the classifier, SGEA approach aims to reduce the injection size and achive the adversarial objectives with minimal perturbation.  
More specifically, it uses deep discriminative subgraph patterns extracted from the CFGs of each class using a correspondence-based quality criterion (CORK) algorithm, which defines a submodular quality criterion that ensures a solution close to the optimal solution~\cite{ThomaCGHKSSYYB10}. This is done by using subgraphs that appear more frequently in one class than others, to fool the machine learning model in predicting that class (\eg when launching a targeted attack). 
Let $D$ denotes the CFGs of the training samples, $D = \{G_i\}^n_{i=1}$ and class labels $C = \{c_i\}^n_{i=1}$ where $c_i \in \{+1,-1\}$ is the class label of graph $G_i$. Also let $D^{+}$ and $D^{-}$ denote the set of graphs in the corresponding classes. For a multi-class dataset, we run the CORK algorithm once for each class where all the graph that belong to the same class are included in $D^{+}$ and the other graphs are in $D^{-}$.
A graph $G_i$ supports another graph $S$ if $S$ is a subgraph of $G_i$. 
Let $D_S=\{G_i| S \subseteq G_i \; \forall \; G_i \in D\}$ denote the supporting graphs of a subgraph $S$.
Moreover, let $D^{+}_{S}$ and $D^{-}_S$, denote the supporting graphs of the subgraph in the positive and negative graphs, respectively.
CORK defines a submodular quality criterion, $q$,
for a subgraph based on the set of supporting graphs (`hits') and non-supporting graphs (`misses') in the two classes and is calculated as follows: 
$q(G_s) = -(|D^{+\sim}_S|*|D^{+\sim}_S| + |D^{+}_S|*|D^{-}_S|)$.
The best quality score is achieved when a subgraph appears in all graphs of one class and not once in the graphs of the other classes; the quality score is 0.
Pruning strategies, as used in the quality criterion of CORK, are integrated into the gSpan algorithm~\cite{YanH02} to directly mine discriminative subgraphs.
Once the set of discriminative subgraphs are mined, we employ gSpan, a graph-based substructure mining pattern for mining frequent subgraphs of size five nodes or higher.

\begin{lstlisting}[float,caption={C script of an example Gafgyt extracted subgraph. Each block is represented as a node in the generated CFG. Appending this code to the source code of a sample will lead to producing the subgraph shown in \autoref{fig:SGEA_Pattern}.},captionpos=b,label={lst:SGEA_Pattern_Script},language=C]
#include<stdio.h>
void  main(){
    int GEAVar1 = 0; // block 0
    if(GEAVar1 == 1){ // block 1 
        GEAVar1 += 1;
    }
    else if(GEAVar1 == 2){ // block 2 
        GEAVar1 += 2;
    }
    int GEAVar2 = 0; // block 3 
    if(GEAVar2 == 0){ // block 4 
        GEAVar2 += 1;
    }
    else{ // block 5 
        GEAVar2 += 2;
    }
    int GEAVar3 = 0; // block 6 
}
\end{lstlisting}

\BfPara{Practical Implementation} The original sample ($x_{org}$) is combined with the selected discriminative subgraph ($x_{sel}$). For example, ~\autoref{fig:SGEA_Pattern} shows the discriminative subgraph extracted from the Gafgyt class and listing~\ref{lst:SGEA_Pattern_Script} shows the equivalent C script to generate that subgraph, which can then be combined with the original sample ($x_{org}$) to generate an AE.
Figure~\ref{fig:PatternExtraction} shows the overview of patterns extraction and the process of generating AEs in the SGEA approach.
While GEA modifies the CFG by connecting the selected graph with the original sample, SGEA connects a carefully generated subgraph with the original sample to generate AE, reducing the injected graph size. 
To generate the subgraph, we extracted the discriminative subgraph patterns from each class, with a size of five nodes or higher. Then, in order to reduce the graph size needed to be injected, we connect the original sample with the subgraph with minimum size. If the generated AE misclassifies, the process succeed, and the AE will be returned; else, we select the next subgraph in ascending order regarding the number of nodes in the subgraph. In case none of the subgraphs cause misclassification, the original sample will be returned as the process failed.

\BfPara{Constructing an AE}
As shown in Figure~\ref{fig:PatternExtraction}, we extract a set of subgraphs from the targeted class. Then, we combine the original sample with the smallest extracted subgraph of the targeted class regarding the number of nodes. If the generated CFG failed to be misclassified by the model, another subgraph is selected in ascending order with respect to the number of nodes in the set of generated subgraphs and combined with the same original graph to generate another CFG.
This process is repeated until a subgraph successfully achieves a misclassification. If no existing subgraph from the set of targeted subgraphs causes misclassification, the original sample is returned, hence generating AE is failed.

\section{\ssmc{}: Design and Evaluation}\label{sec:No_defense}
This section presents \ssmc{}, \textbf{D}eep \textbf{L}earning-based \textbf{S}ingle \textbf{S}hot \textbf{M}alware \textbf{C}lassification approach. We describe the design and methods for \ssmc{} in~\textsection\ref{sec:NDsystemDes} and present the evaluation in~\textsection\ref{sec:Results}.

\begin{figure}[t]
\centering
\includegraphics[width=0.49\textwidth]{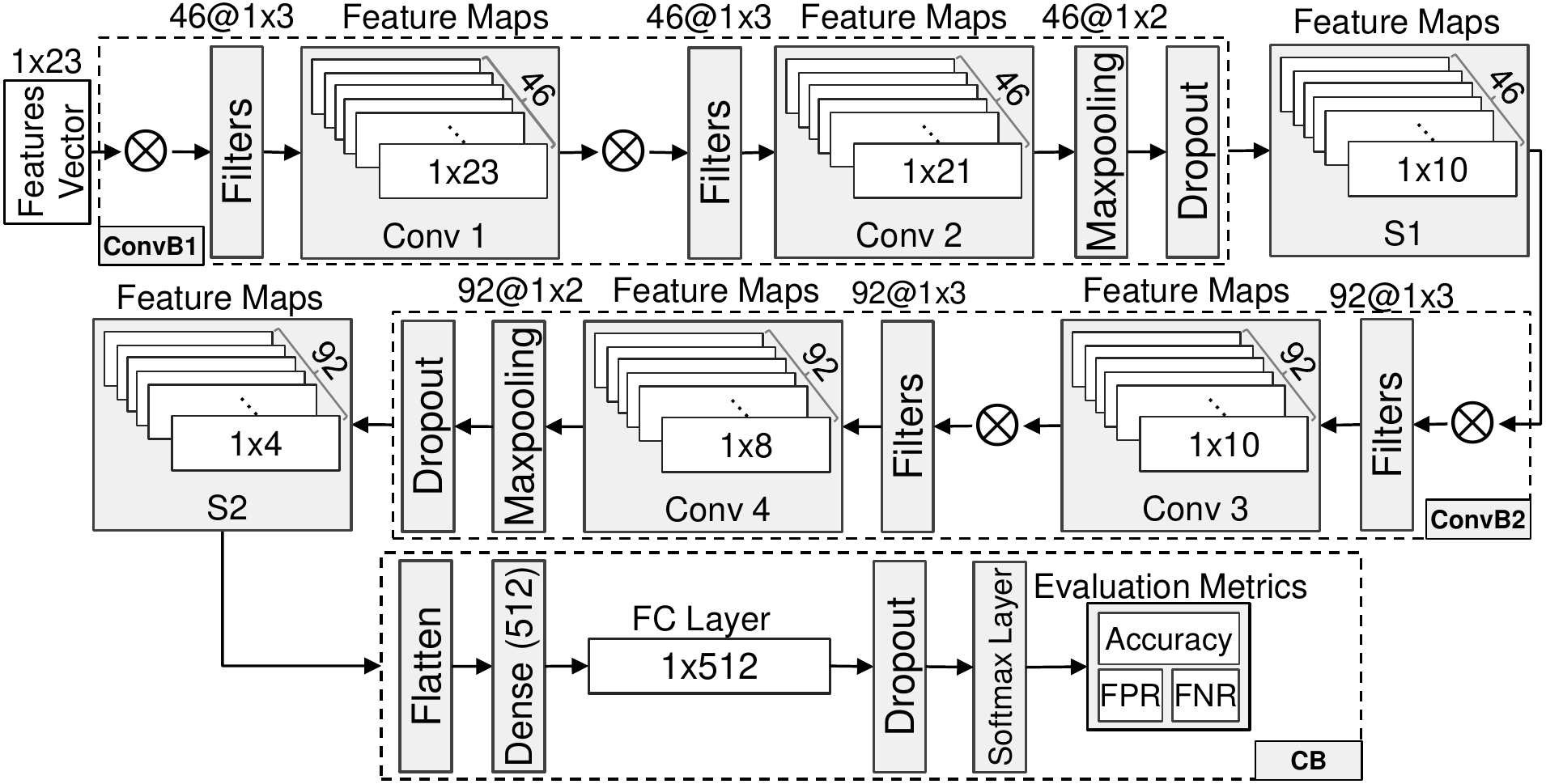}
\caption{Internal design of the CNN architecture used for detection and classification task. Notice that 46@1x3, for example, refers to applying 46 filters each of size 1x3 on the input data. The design consists of four convolutional layers with maxpooling and dropout operations. Then, a dense layer of size 512 is used with a softmax activation function to output the prediction.}
\label{fig:CNN_Architecture}
\end{figure}

\subsection{\ssmc{}: System Design} \label{sec:NDsystemDes}

For the malware detection design, \ssmc{} aims to distinguish the malicious IoT applications from the benign. The implementation of \ssmc{} incorporates deep learning models, \eg Convolutional Neural Network (CNN)-based and Deep Neural Network (DNN)-based models, trained using the extracted CFG-based algorithmic features. 
The input ($X$) to the model is a one-dimensional (1D) vector of size $1 \times 23$ representing the extracted features.
We explore the capabilities of two deep learning architectures, namely CNN and DNN. We describe the separate designs in the following.

\BfPara{CNN-based Design} The CNN-based \ssmc{} model constitutes of three blocks, namely convolutional block 1 (ConvB1), convolutional block 2 (ConvB2), and classification block (CB). \autoref{fig:CNN_Architecture} illustrates the CNN-based \ssmc{} model architecture.
\begin{itemize}[leftmargin=1.5em]
    \item \BfPara{ConvB1} This block takes in feature vector ($X$) as the input. Architecturally, it is made up of two convolutional layers, a 1D convolutional layer (Conv 1) with padding and 46 filters (${F_{b1}}'$) of size $1 \times 3$, convolving over the input data ($X$) with stride of 1, resulting in a 2D tensor (${C_{b1}}'$) of size $1 \times 23$. Followed by a similar 1D convolutional layer (Conv 2) without padding. The output of this layer is a 2D tensor (${C_{b1}}''$) of size $46 \times 21$. Then, a maxpooling operation of size and stride of 2 and dropout with probability of 0.25 are applied. The output of this block is a 2D tensor ${S_{b1}}$ of size $46 \times 10$.
       \[ {C_{b1}}'_i=X \otimes {F_{b1}}'_i\] 
        \[{C_{b1}}''_i={C_{b1}}'_i \otimes {F_{b1}}''_i \] 
        \[{M_{b1}}_i=\text{\sf maxpool}({C_{b1}}''_i, 2, 2)\] 
        \[{S_{b1}}_i=\text{\sf dropout}({M_{b1}}_i, 0.25)\] 
    
    \item \BfPara{ConvB2} The result of the ConvB1 block (${S_{b1}}$) is fed into this block. It is similar to ConvB1, except for the difference in the number of filters (${F_{b2}}'$) in the convolutional layers. This block consists of a 1D convolutional layer (Conv 3) with padding and 92 filters of size $1 \times 3$, convolving with a stride of 1. Followed by a similar 1D convolutional layer (Conv 4) without padding. The output of this layer is a 2D tensor ${C_{b2}}''$ of size $92 \times 8$. Maxpooling operation of size and stride of 2 is applied, followed by a dropout with a probability of 0.25, resulting in a 2D tensor ${S_{b2}}$ of size $92 \times 4$.
       \[{C_{b2}}'_i={S_{b1}} \otimes {F_{b2}}'_i\] 
       \[{C_{b2}}''_i={C_b2}'_i \otimes {F_{b2}}''_i\] 
       \[{M_{b2}}_i=\text{\sf maxpool}({C_{b2}}''_i, 2, 2)\] 
       \[{S_{b2}}_i=\text{\sf dropout}({M_{b2}}_i, 0.25)\] 
    
    \item \BfPara{CB} The generated tensor in ConvB2 (${S_{b2}}$) is then fed into this block. The input is forwarded to flatten operation, resulting in a 1D tensor ($F_{l}$) of size $1 \times 368$. Followed by a fully connected (FC) dense layer (${FCL}$) of size 512 and a dropout with a probability of 0.5 resulting in ${S_{FC}}$. Finally, ${S_{FC}}$ is fed to the softmax layer as the output layer. 
       \[{F_{l}} = \text{Flatten}(S_{b2})\]
       \[{FCL}=\text{\sf dense}(F_{l},512)\]
       \[{S_{FC}}=\text{\sf dropout}({FCL}, 0.5)\]
       \[\text{output}=\text{\sf softmax}({S_{FC}})\]%
\end{itemize}

\begin{figure}[t]
\centering
\includegraphics[width=0.49\textwidth]{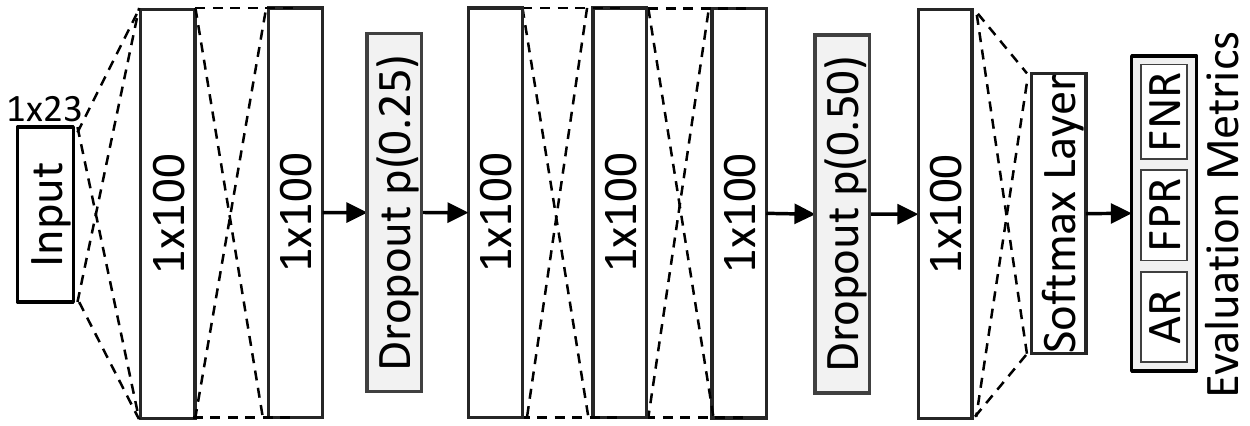}
\caption{The internal architectural of the DNN used for the detection and classification tasks. The design consists of six fully connected layers, with dropout operations and softmax activation function.}
\label{fig:DNN_Architecture}
\end{figure}

\BfPara{DNN-based Design} \autoref{fig:DNN_Architecture} illustrates the DNN-based \ssmc{} model. The architecture of the DNN-based model consists of two consecutive and fully connected dense layers of size $1\times100$ connected to the input vector, followed by a dropout operation with a probability of 0.25. The output of the dropout function is fed to fully connected with another two fully connected dense layers of size $1\times100$, succeeded by a dropout operation with a probability of 0.5. The output is then fed to the softmax layer. 
This design enables the extraction of deep feature representations and patterns from the feature vectors, and therefore, finding discriminative characteristics for detection and classification processes.

\BfPara{Training Process}
We trained the two model architectures, \ie CNN-based and DNN-based \ssmc{}, using 100 epochs with a batch size of 32. Both the convolutional and the fully connected layers use Rectified Linear Units (ReLU) as the activation function. For regularization, we use dropout to prevent model over-fitting and allow the propagation of robust and distinct features through the model layers. 

\BfPara{Malware Classification Task}
Malware classification task refers to identifying the malware family of targeted samples, while the detection process aims to detect malicious behavior regardless of the category to which the malicious behaviors belong. Therefore, the detection task can be viewed as a binary classification task that aims to detect malicious behavior from the extracted features of the CFG. The classification task, on the other hand, aims to detect and identify the family of the malicious behavior.

\BfPara{Experimental Setup}
We evaluate \ssmc{} using the dataset in~\textsection\ref{sec:methodology}, and show the robustness against AEs generated using GEA and SGEA. 
The experiments were conducted on a workstation operating Ubuntu 16.04 and comprising of an i5-8500 CPU, 32GB DDR4 RAM, 512GB SSD and 4TB HHD of storage, and NVIDIA RTX2080 Ti Graphics Processing Unit (GPU). The experiments are implemented and conducted using Python 3.5.2. 

\BfPara{Evaluation Metrics} The performance of the system is reported using the following evaluation metrics: 1) The accuracy of the model, computed as the ratio of the correctly labeled samples ($CLS$) over all test samples ($|D|$), defined as: ${CLS} \div |D|$. 2) False Positive Rate (FPR), which is the number of incorrectly labeled malware samples ($ILM$) over the total number of malicious samples ($|D_m|$), computed as $ILM \div |D_m|$. 3) False Negative Rate (FNR), represented as the incorrectly labeled benign samples ($ILB$) divided by the total number of benign samples ($|D_b|$), $ILM \div |D_b|$. Finally, the confusion matrix, where the columns represent the actual classes and the rows are the predicted labels. The value at a location ($x$,$y$) represent the portion of the samples of class $x$ classified as $y$.

\subsection{\ssmc{}: Evaluation and Results} \label{sec:Results}

\subsubsection{\ssmc{}: Baseline Performance}\label{sec:Res_DLMDS}

\BfPara{\ssmc{}: Detection Task}
We designed two-class detection \ssmc{} using the CNN- and DNN-based models that distinguish IoT malware from the IoT benign applications. The model is trained over 23 CFG-based graph-theoretic features categorized into seven groups. 
The two models, CNN- and DNN-based, achieve an accuracy rate of 98.96\% and 98.67\% with a FNR of 0.88\% and 1.41\% and FPR of 1.33\% and 1.16\%, respectively.~\autoref{DetectionSystems} shows the evaluation of each trained model.

\begin{table}[t]
\centering
\caption{Evaluation of the CNN- and DNN-based IoT malware detection systems on normal samples (\ie non-adversarial). Here, the CNN-based model outperforms its counterpart in both accuracy and FNR, achieving a detection accuracy of 98.96\%.}
\label{DetectionSystems}
\begin{tabular}{c|c|c|c|c}
\toprule
Architecture & Accuracy & FNR & FPR & Time/Epoch\\
\midrule
CNN & 98.96\% & 0.88\% & 1.33\% & 1.16s\\
DNN & 98.67\% & 1.41\% & 1.16\% & 0.21s\\
\bottomrule
\end{tabular}
\end{table}

\begin{table}[t]
\centering
\caption{Evaluation of the CNN- and DNN-based IoT malware classification systems on normal samples (\ie non-adversarial). Similar to \autoref{DetectionSystems}, the CNN-based model outperforms the DNN-based model, achieving a classification accuracy of 98.09\%. However, the training time is higher, 1.23 seconds per epoch, in compare to 0.22 seconds for DNN-based model.}
\label{ClassificationSystems}
\begin{tabular}{c|c|c}
\toprule
Architecture & Accuracy & Time/Epoch\\
\midrule
CNN & 98.09\% & 1.23s\\
DNN & 97.57\% & 0.22s\\
\bottomrule
\end{tabular}
\end{table}

\begin{figure}[t]
    \centering
    \begin{subfigure}[b]{0.23\textwidth}
        \centering
        \includegraphics[width=0.95\linewidth]{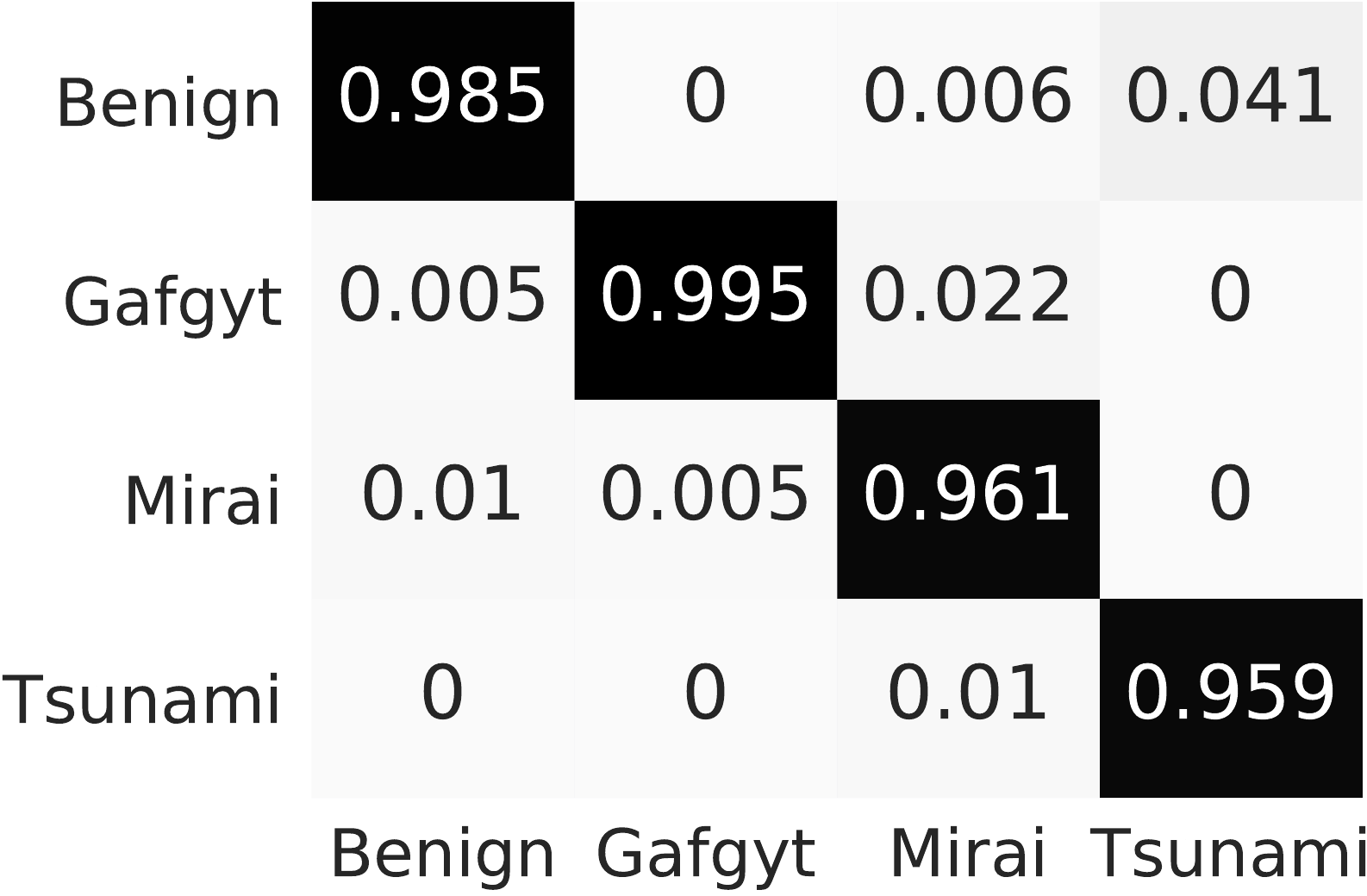}
        \caption{CNN-based}
        \label{fig:ClassificationEvalCNN}
    \end{subfigure}
    \begin{subfigure}[b]{0.23\textwidth}
        \centering
        \includegraphics[width=0.95\linewidth]{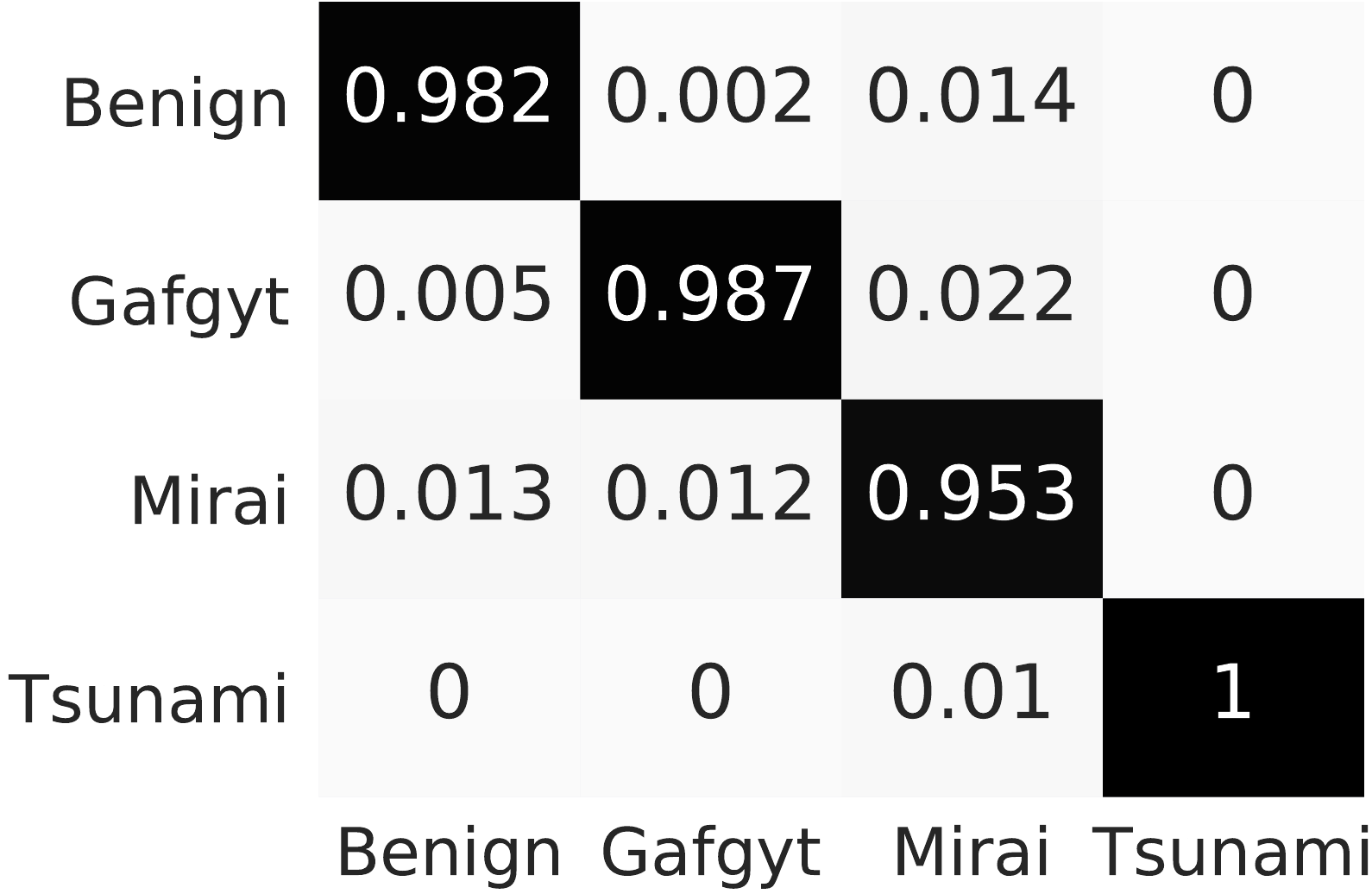}
        \caption{DNN-based}
        \label{fig:ClassificationEvalDNN}
    \end{subfigure}%
    \caption{Confusion matrices of IoT malware classification systems. Here, each column represents the actual class, whereas, rows represents the predicted labels.}
    \label{fig:ClassificationEval}
\end{figure}

\BfPara{\ssmc{}: Classification Task}
In addition to detecting the IoT malicious samples, we also design a four-class classification \ssmc{} incorporating CNN- and DNN-based model designs. The classification task aims to evaluate \ssmc{} for classifying the malicious samples into their corresponding families. We achieved a CNN- and DNN-based model accuracy rate of 98.09\% and 97.57\%, respectively as shown in~\autoref{ClassificationSystems}. We also provide the the confusion matrices in ~\autoref{fig:ClassificationEvalCNN} and~\autoref{fig:ClassificationEvalDNN} to report the performance of CNN-based and DNN-based \ssmc{} for the individual class as each column represents the ground truth class, and every row represents the predicted labels.

    \begin{figure*}[htb]
        \centering
        \begin{subfigure}[b]{0.24\textwidth}
            \centering
            \includegraphics[width=0.95\linewidth]{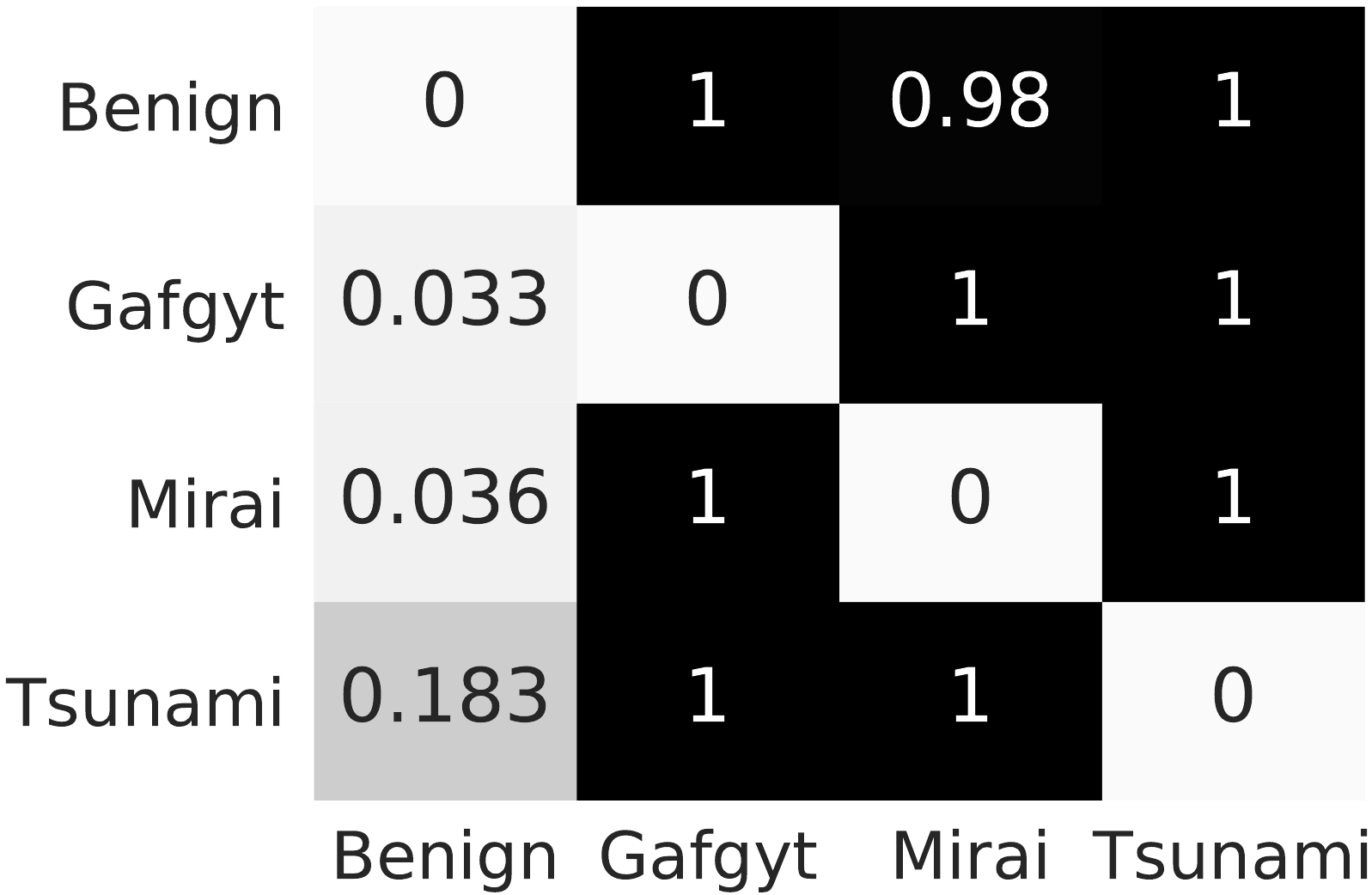}
            \caption{Non-targeted MR}
            \label{fig:ClassificationResCNNU}
        \end{subfigure}
        \begin{subfigure}[b]{0.24\textwidth}
            \centering
            \includegraphics[width=0.95\linewidth]{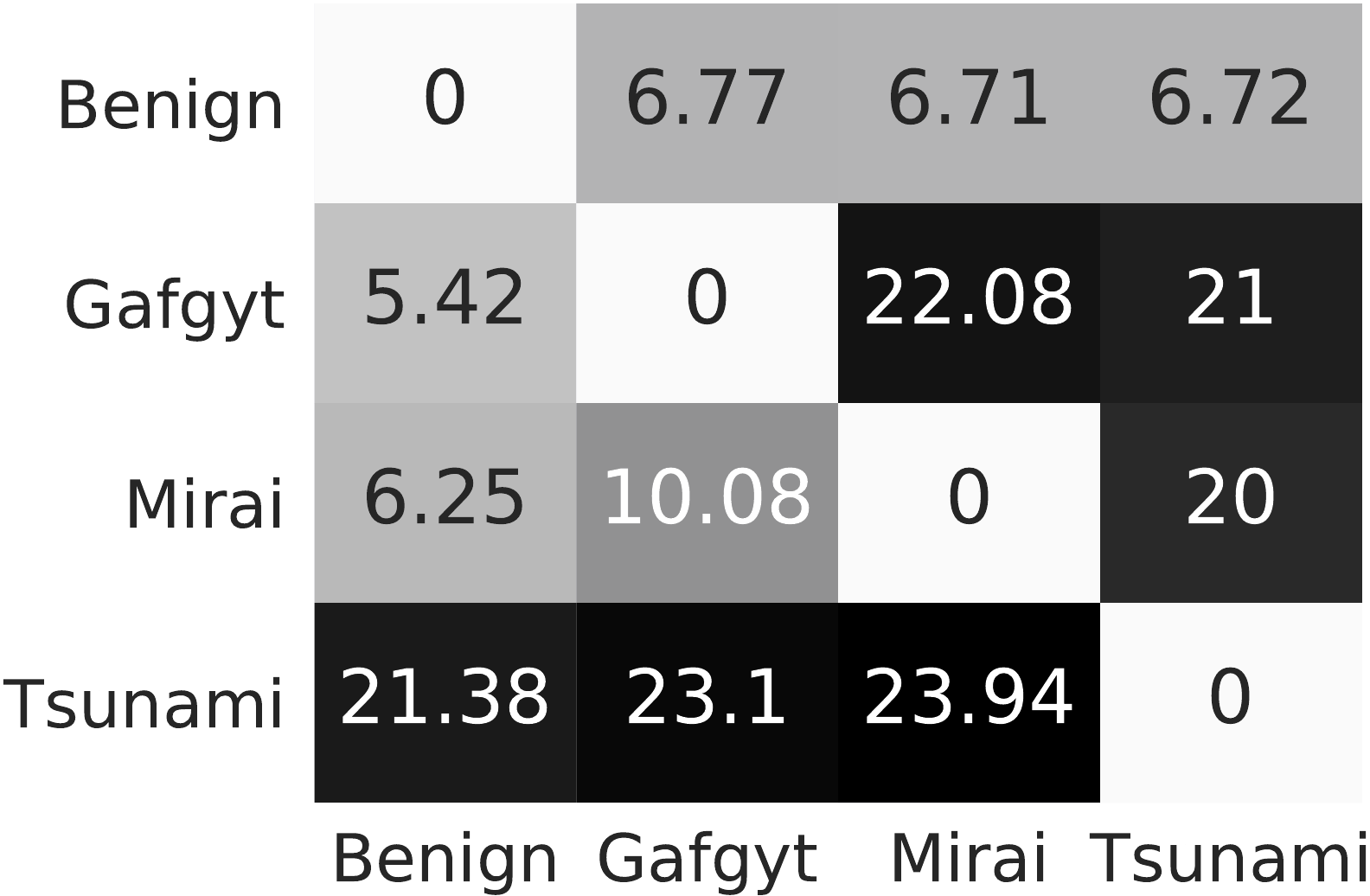}
            \caption{Non-targeted MR subgraph size}
            \label{fig:ClassificationResCNNUNodes}
        \end{subfigure}
        \begin{subfigure}[b]{0.24\textwidth}
            \centering
            \includegraphics[width=0.95\linewidth]{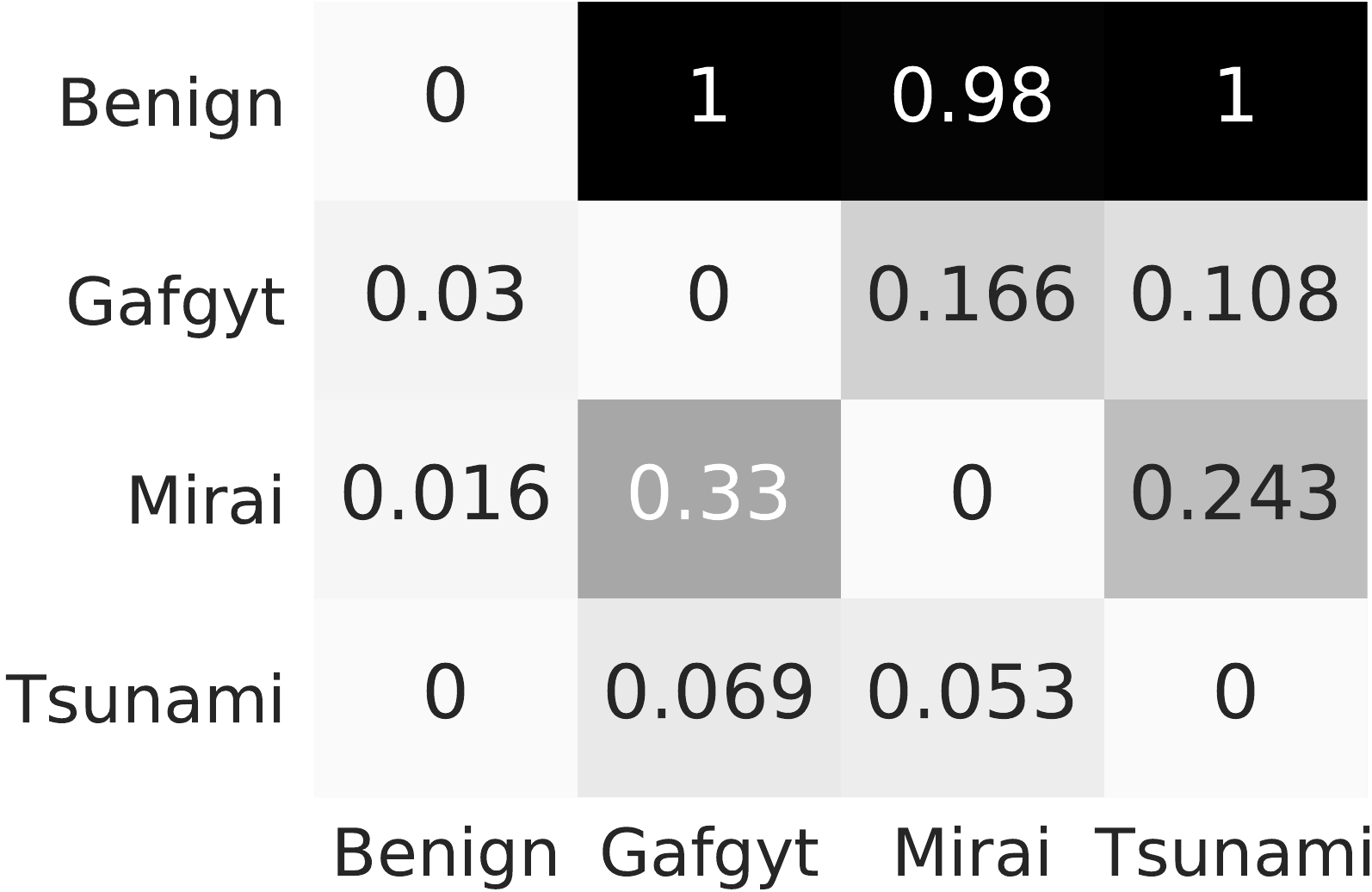}
            \caption{Targeted MR}
            \label{fig:ClassificationResCNNT}
        \end{subfigure}
        \begin{subfigure}[b]{0.24\textwidth}
            \centering
            \includegraphics[width=0.95\linewidth]{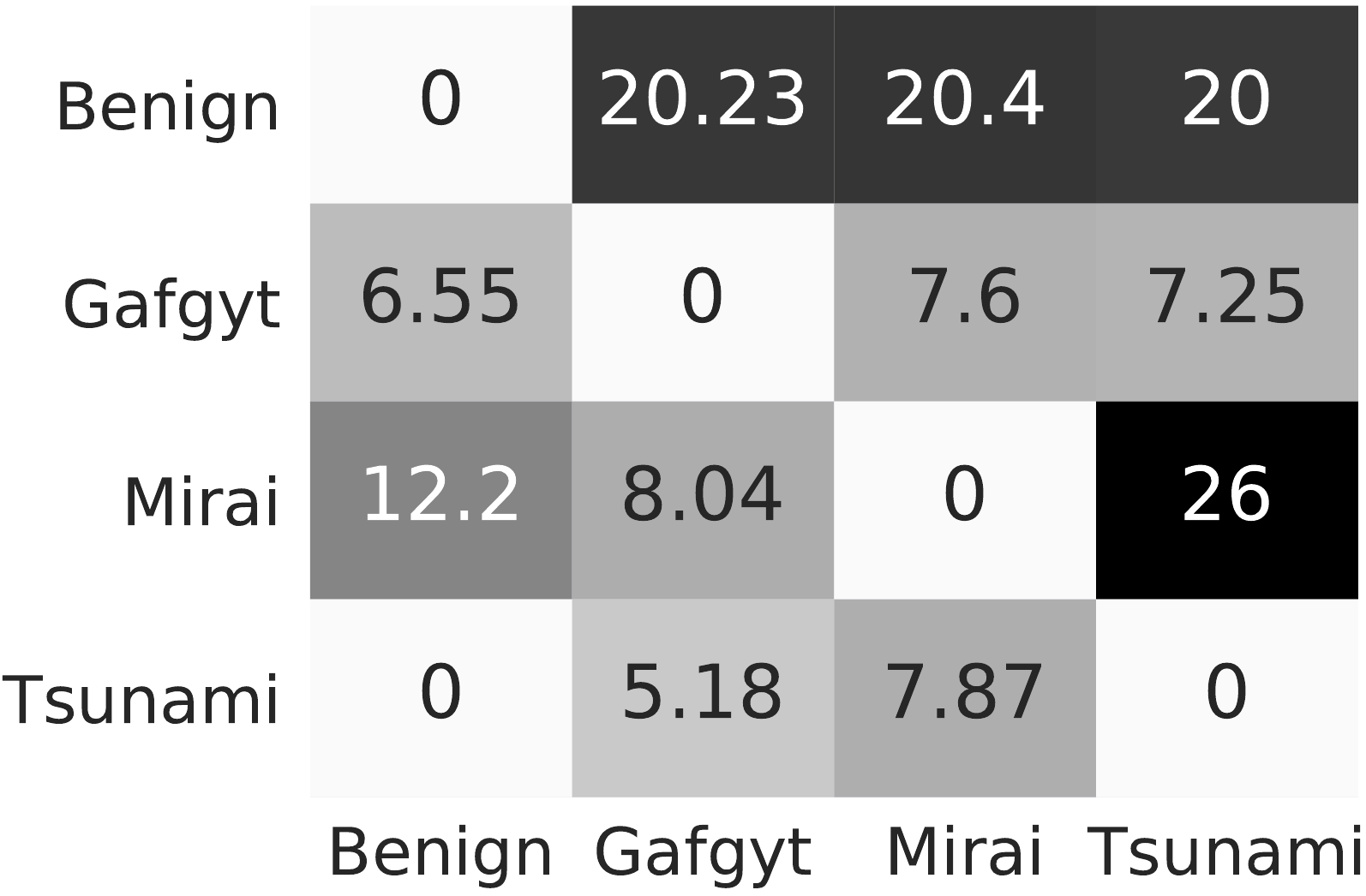}
            \caption{Targeted MR subgraph size}
            \label{fig:ClassificationResCNNTNodes}
        \end{subfigure}
        \caption{SGEA: CNN-based IoT malware classification system evaluation. Here, MR refers to misclassification rate, columns represent the sample original class, whereas, rows represent the connected subgraph pattern class.}
        \label{fig:ClassificationResCNN}
    \end{figure*}
    
    \begin{figure*}[htb]
        \centering
        \begin{subfigure}[b]{0.24\textwidth}
            \centering
            \includegraphics[width=0.95\linewidth]{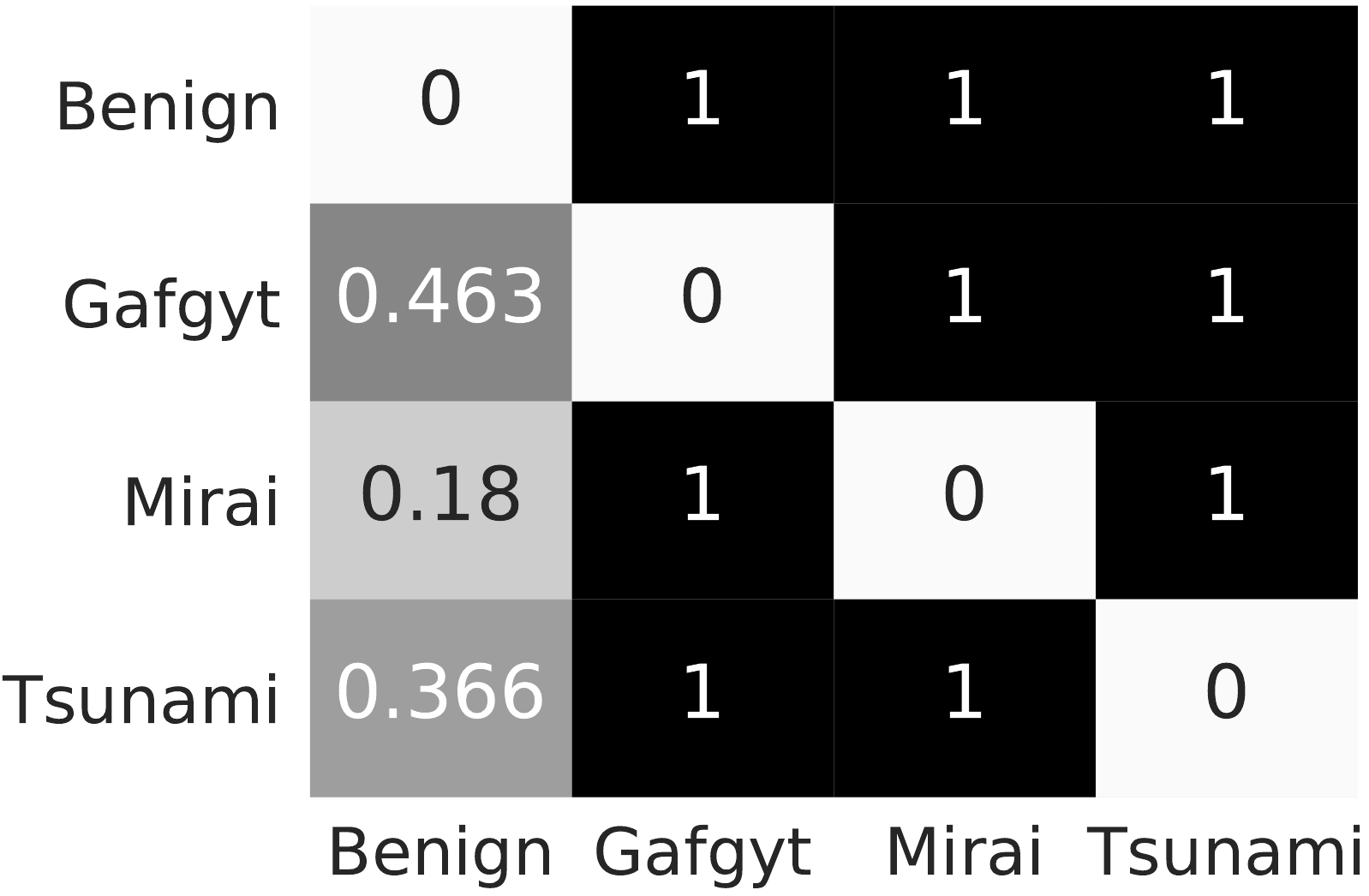}
            \caption{Non-targeted MR}
            \label{fig:ClassificationResDNNU}
        \end{subfigure}
        \begin{subfigure}[b]{0.24\textwidth}
            \centering
            \includegraphics[width=0.95\linewidth]{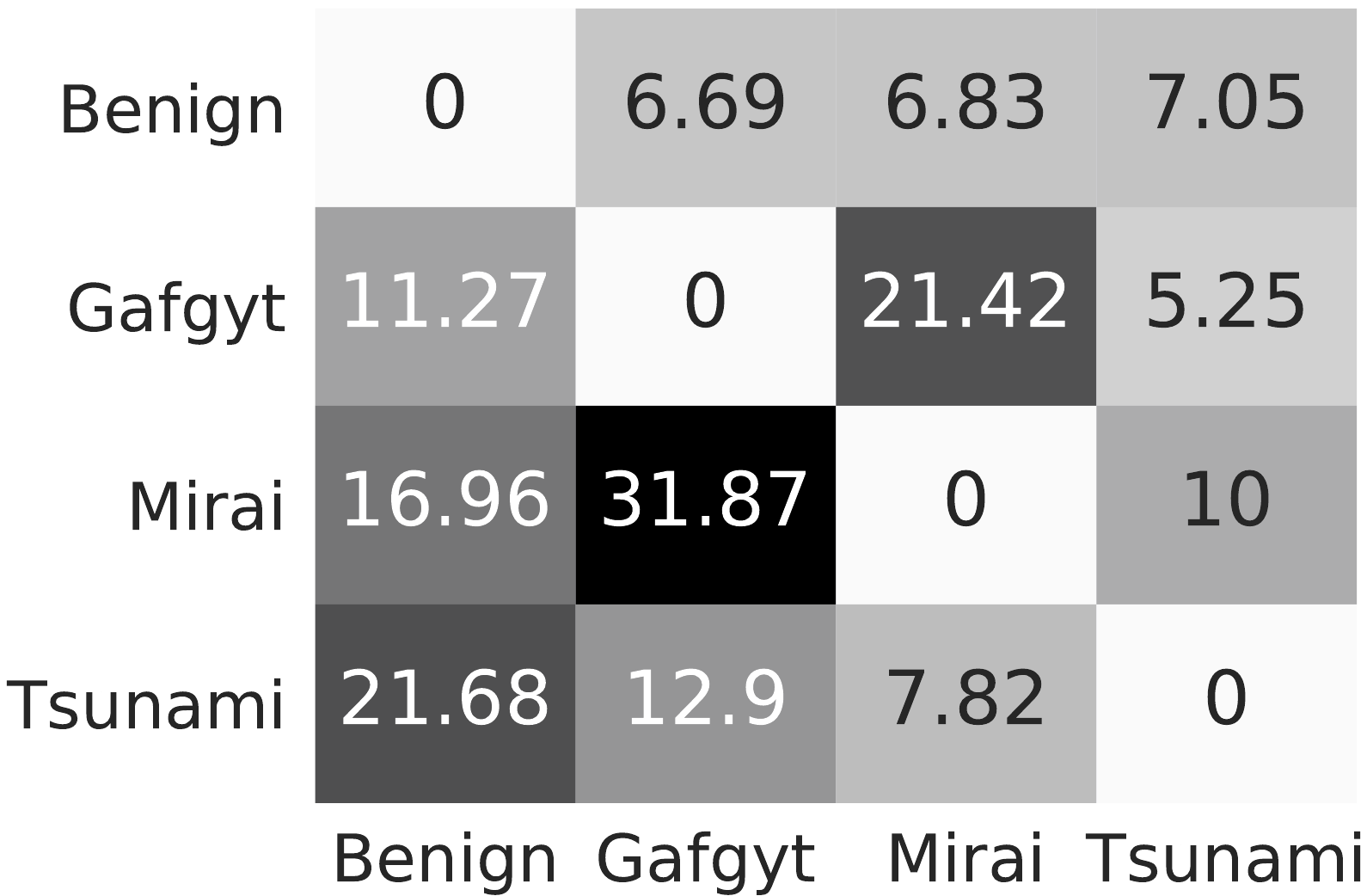}
            \caption{Non-targeted MR subgraph size}
            \label{fig:ClassificationResDNNUNodes}
        \end{subfigure}
        \begin{subfigure}[b]{0.24\textwidth}
            \centering
            \includegraphics[width=0.95\linewidth]{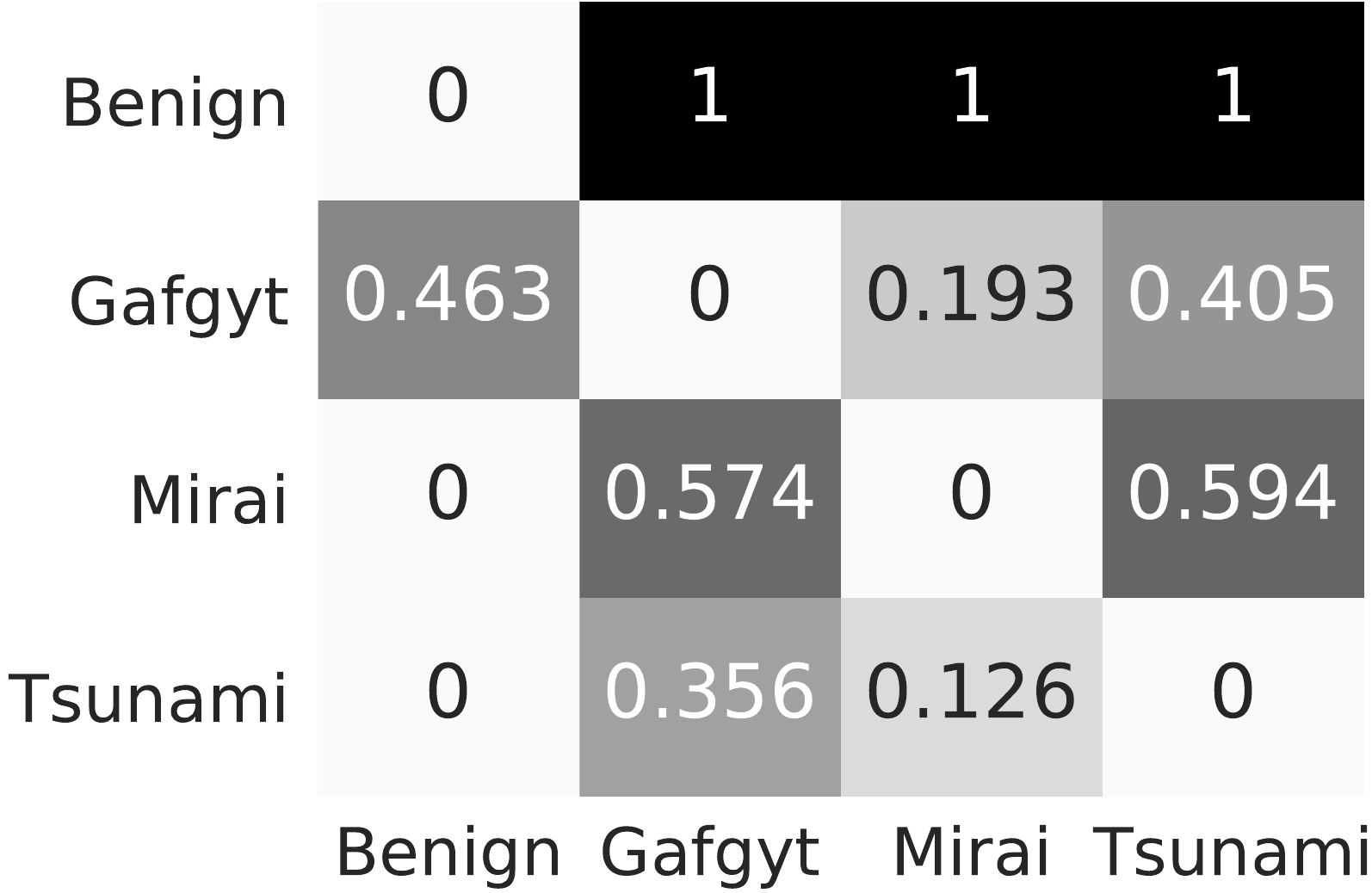}
            \caption{Targeted MR}
            \label{fig:ClassificationResDNNT}
        \end{subfigure}
        \begin{subfigure}[b]{0.24\textwidth}
            \centering
            \includegraphics[width=0.95\linewidth]{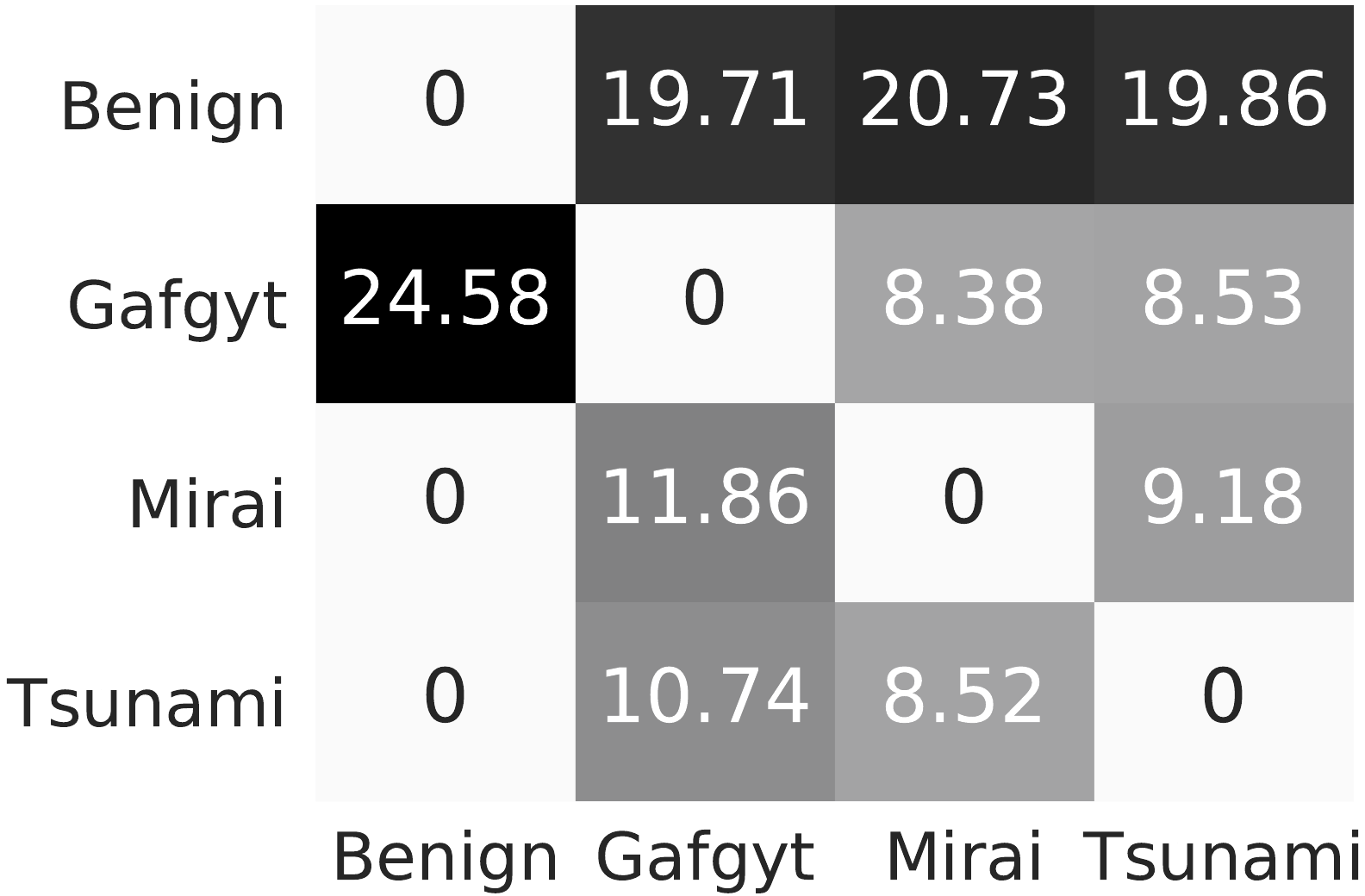}
            \caption{Targeted MR subgraph size}
            \label{fig:ClassificationResDNNTNodes}
        \end{subfigure}
        \caption{SGEA: DNN-based IoT malware classification system evaluation. Here, MR refers to misclassification rate, columns represent the sample original class, whereas, rows represent the connected subgraph pattern class.}
        \label{fig:ClassificationResDNN}
    \end{figure*}

\subsubsection{\ssmc{}: Robustness Assessment against GEA}\label{sec:Res_GC}
We investigate the robustness of \ssmc{} against AEs generated using GEA.
GEA is designed to generate practical AEs that fools the classifier while preserving the functionality and executability of the original sample. 
We  explore the impact of the size of the graph and discuss the fundamental overhead of using GEA. 
Note that all generated samples maintain the practicality and functionality of the original code. 
From the benign and malicious IoT software, we selected three graphs as targets (the selected samples have the minimum, median, and maximum size across the dataset). and connected each of these target graphs with every graph in the testing dataset of the other class. 
We used the target graphs to have different sizes, \eg minimum, median, and maximum graph size across the dataset, to understand the impact of size on the misclassification rate introduced by AEs using GEA. 

\BfPara{Robustness of Detection Models}
The results of \ssmc{} performance against AEs generated by GEA are shown in~\autoref{GCNodesMB} and~\autoref{GCNodesBM}. 
Intuitively, a key finding is the impact of graph size on the misclassification rate, since the increase in the graph size, \ie the included number of nodes, results in higher misclassification rate. However, this success rate of adversarial objectives comes with a computational cost since the reported time for crafting AEs is proportional to the size of the injected graph as perturbation. 
In the IoT malware detection task, we achieved a malware to benign misclassification rate of as high as 100\% on both CNN- and DNN-based models, and a benign to malware misclassification rate of 60.22\% and 60.89\% on the respective models. 

\BfPara{Robustness of Classification Models}
\autoref{GCNodesClassification} shows the targeted and non-targeted misclassification rates over the IoT malware classification task. Non-targeted misclassification indicates that after combining the original and selected samples, the original sample class changes. On the other hand, if the new assigned label is the selected sample class, it is considered as targeted misclassification. 
Here, we achieved a targeted misclassification rate of 100\% from all malicious families, \ie when designing AEs with benign-selected samples so the models misclassify them to benign, using both the CNN- and DNN-based models. 
Similar to the IoT malware detection system, the targeted and non-targeted misclassification rates increase with the increase in the number of nodes for the injected graph. 

\begin{table}[t]
\centering
\caption{GEA: Malware to benign misclassification rate over IoT detection systems. MR refers to misclassification rate, whereas, CT refers to crafting time in millisecond per sample.}
\label{GCNodesMB}
\begin{tabular}{c|c|c|c|c}
\toprule
\multirow{2}{*}{Size} & \multirow{2}{*}{\# Nodes} & \multicolumn{2}{c|}{MR (\%)} & \multirow{2}{*}{CT (ms)} \\
&  & CNN & DNN &  \\

\midrule

Minimum &  10 & 50.57 & 61.03 & 37.39  \\ 
Median &  23 & 99.64 & 98.76 & 40.46  \\ 
Maximum &  1075 & 100 & 100 & 6,430.66 \\ 

\bottomrule
\end{tabular}
\end{table}

\begin{table}[t]
\centering
\caption{GEA: Benign to malware misclassification rate over IoT detection systems. MR refers to misclassification rate, whereas, CT refers to crafting time in millisecond per sample.}
\label{GCNodesBM}
\begin{tabular}{c|c|c|c|c}
\toprule
\multirow{2}{*}{Size} & \multirow{2}{*}{\# Nodes} & \multicolumn{2}{c|}{MR (\%)} & \multirow{2}{*}{CT (ms)} \\
&  & CNN & DNN &  \\

\midrule

Minimum &  11 & 45.92 & 19.33 & 34.10  \\ 
Median &  43 & 60.22 & 59.90 & 56.83  \\ 
Maximum &  274 & 47.36 & 60.89 & 763.63 \\ 

\bottomrule
\end{tabular}
\end{table}

\begin{table}[t]
\centering
\caption{GEA: Misclassification rate over IoT classification systems. MR and TMR refers to non-targeted and targeted misclassification rates, respectively, whereas, CT refers to the crafting time in millisecond per sample.}
\label{GCNodesClassification}
\begin{tabular}{c|c|c|c|c|c|c}
\toprule
\multirow{2}{*}{Class} & \multirow{2}{*}{\# Nodes} & \multicolumn{2}{c|}{MR (\%)} & \multicolumn{2}{c|}{TMR (\%)} & \multirow{2}{*}{CT (ms)} \\
 & & CNN & DNN & CNN & DNN &  \\

\midrule
\multirow{2}{*}{Benign} & 10 & 48.72 & 58.87 & 45.89  & 48.54  &  37.39  \\ 
                        & 23 & 99.64 & 99.55 & 99.64  & 99.38  & 40.46  \\ 
                        & 1075 & 100 & 100 & 100  & 100  & 6,430.66 \\ 
\midrule

\multirow{2}{*}{Gafgyt} & 13 & 24.62 & 41.74 & 0.17  & 0.35  & 32.36 \\ 
                        & 64 & 66.81 & 77.40 & 15.97  & 16.06  & 68.77  \\ 
                        & 155 & 54.63 & 47.13 & 8.03  & 0.00  & 125.15 \\ 
\midrule

\multirow{2}{*}{Mirai} & 11 & 41.37 & 37.22 &  0.32 &  0.80 & 34.10  \\ 
                        & 48 & 62.30 & 52.15 &  12.46 &  0.96 & 56.84  \\ 
                        & 274 & 95.60 & 91.05 & 93.45  & 53.67  & 763.63 \\ 
\midrule

\multirow{2}{*}{Tsunami} & 15 & 59.54 & 60.73 & 0.12  & 0.12  & 35.25  \\ 
                        & 59 & 63.95 & 64.06 &  0.00 &  0.00 & 59.93  \\ 
                        & 138 & 66.74 & 64.36 &  0.00 & 0.00  & 201.82 \\ 

\bottomrule
\end{tabular}
\end{table}

\subsubsection{\ssmc{}: Robustness Assessment against SGEA}\label{sec:Res_GCOPT}
The GEA achieves high misclassification rate by injecting the original sample with another selected sample from a target class to fool the models in predicting the wrong class. The GEA approach comes with computational cost and increased binary size that accommodate the combination of two samples into one.
These costs are reduced by using 
the SGEA approach that reduces the size of injection to carefully selected subgraph that achieves the adversarial objective.
    
\BfPara{Robustness of Detection Models}
\autoref{SGEADetectionB2M} and \autoref{SGEADetectionM2B} show the results of SGEA against CNN- and DNN-based \ssmc{} detection models. 
Notice that GEA outperforms SGEA in benign to malware misclassification. However, SGEA achieves 100\% malware to benign misclassification rate against CNN- and DNN-based models, outperforming the GEA approach with an average subgraph size of 6.80 and 6.86, respectively.

    \begin{table}[t]
        \centering
        \caption{SGEA: Benign to malware IoT malware detection system evaluation. Here, MR refers to misclassification rate, AVG. Size refers to the overall average subgraph size used to achieve misclassification, and CT is  sample's crafting time in seconds.}
        \label{SGEADetectionB2M}
        \begin{tabular}{c|c|c|c}
            \toprule
            Architecture & MR(\%) & AVG. Size & CT (s) \\
            
            \midrule
            
            CNN &  22.22 & 10.15 & 2.57  \\ 
            DNN &  33.88 & 11.09 & 2.23   \\
            
            \bottomrule
        \end{tabular}
    \end{table}
    \begin{table}[t]
        \centering
        \caption{SGEA: Malware to benign IoT malware detection system evaluation. Here, MR refers to misclassification rate, AVG. Size refers to the overall average subgraph size used to achieve misclassification, and CT is sample's crafting time in seconds.}
        \label{SGEADetectionM2B}
        \begin{tabular}{c|c|c|c}
            \toprule
            Architecture & MR(\%) & AVG. Size & CT (s) \\
            
            \midrule
            
            CNN &  100 & 6.80 & 0.23  \\ 
            DNN &  100 & 6.86 & 0.21   \\
            
            \bottomrule
        \end{tabular}
    \end{table}
    
\BfPara{Robustness of Classification Models}
\autoref{fig:ClassificationResCNN} and~\autoref{fig:ClassificationResDNN} show the results of the CNN- and DNN-based \ssmc{} classification models against the SGEA approach. \autoref{fig:ClassificationResCNNU} and \autoref{fig:ClassificationResCNNT} represent the non-targeted and targeted misclassification rate using the CNN-based model, respectively. Similarly,~\autoref{fig:ClassificationResDNNU}, and~\autoref{fig:ClassificationResDNNT} show the non-targeted and targeted misclassification rate using the DNN-based model, respectively. 
While the columns represent the sample's original class, the rows represent the class that subgraph pattern was extracted from. Targeted misclassification occurs when the AE's predicted label is not different than the class that the subgraph pattern belongs to.
~\autoref{fig:ClassificationResCNNUNodes},~\autoref{fig:ClassificationResCNNTNodes},~\autoref{fig:ClassificationResDNNUNodes} and~\autoref{fig:ClassificationResDNNTNodes} represent the average size of the connected subgraphs to generate the AEs over the CNN- and DNN-based models. 
For instance, SGEA approach successfully misclassifies all Gafgyt test samples into benign using the CNN-based classification model with an average subgraph size of 20.23 nodes. 
Moreover, it misclassifies all Gafgyt test samples into other classes using discriminative subgraphs from benign samples with an average size of 6.77 nodes. 
Note that all classes have a high targeted misclassification rate toward benign class and low targeted misclassification rate from benign to malicious families and among the malicious families. This is a result of the nature of the benign samples; they are diverse in characteristics and functionalities. However, malicious samples within the same family are more likely to share similar behaviors, resulting in distinctive traits that can be captured from the extracted CFG.

\BfPara{AE Crafting Time (GEA vs. SGEA)} 
The SGEA significantly reduces the size of AEs and introduces more sophisticated adversarial settings by incorporating reduced perturbation to the software sample.
However, similar to GEA, the SGEA approach is computationally extensive. We observe an average AE crafting time of 4.74 and 3.95 seconds per sample using CNN- and DNN-based classification models, in comparison to the average crafting time of 0.65 seconds for the GEA approach.


\section{\fhmc{}: Coping with AEs}\label{sec:sbd}
In this section, we propose and describe the \fhmc{} approach, \textbf{F}ine-grained \textbf{H}ierarchical Learning for \textbf{M}alware \textbf{C}lassification.
Machine learning methods for malware detection and classification, \eg \ssmc{}, are susceptible to AEs and fall short of delivering a  robust system against adversarial settings, as shown in ~\textsection\ref{sec:No_defense}. This motivates to explore methods and alternative designs to cope with such vulnerabilities to adversarial attacks. We propose \fhmc{}, a robust system for malware detection and classification that leverages deep learning on 
a fine-grained and hierarchical manner to detect suspicious malicious behaviors.

\begin{figure*}[t]
\centering
\includegraphics[width=0.9\textwidth]{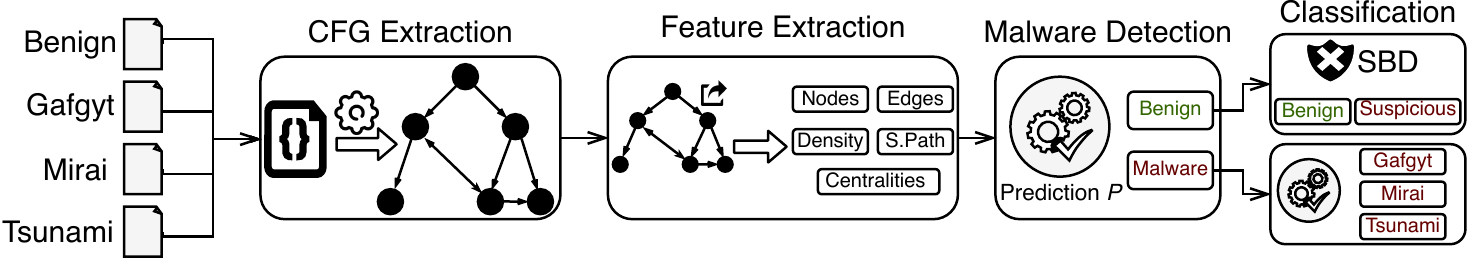}
\caption{Fine-grained hierarchical learning classification flow. First, corresponding CFGs of the IoT software are generated, then, 23 algorithmic features are extracted from the CFGs. Afterward, an IoT malware detection system classifies samples into benign and malware, all malware samples are directed to IoT malware classification system, while benign samples are directed into suspicious behavior detection system for further investigation. SBD refers to suspicious behavior detection system.}
\label{fig:SBD_flow}
\end{figure*}

\subsection{\fhmc{}: System Design}
The design of \fhmc{} consists of five components as illustrated in~\autoref{fig:SBD_flow}. Description of each component is in the following.

\begin{itemize}[leftmargin=1.5em]
    \item \BfPara{CFG Extraction} This component is responsible for extracting the CFGs of the software samples using Radare2, and presenting them as labeled CFGs for further analysis.
    
    \item \BfPara{Feature Extraction} The feature extraction component calculates 23 algorithmic features from the samples' CFGs. Details of the extracted features are in~\textsection\ref{sec:Dataset} and~\autoref{DSFeatures}.
    
    \item \BfPara{Malware Detection} The malware detection component utilizes a CNN-based model that has the model architecture of~\autoref{fig:CNN_Architecture}. The purpose of this model is to classify the samples into malware and benign. Samples classified as benign are directed to a suspicious behavior detection process. Samples classified as malware are directed to a classification model.
    
    \item \BfPara{Malware Classification} This component is fed by the samples classified as malware by the malware detection component. The goal of this component is to classify the sample into three IoT malicious families, \ie Gafgyt, Mirai, and Tsunami. The design and architecture of this component is similar to the CNN-based classification model of~\autoref{fig:CNN_Architecture}.
    
    \item \BfPara{Suspicious Behavior Detector} This component detects a potential suspicious behavior within the samples. Since the adversary may generate AEs with the purpose of fooling the system in assigning them to benign class, this component further investigates the potential of suspicious behavior within the benign-classified sample using the extracted CFG. 
\end{itemize}

\begin{figure*}[t]
\centering
\includegraphics[width=0.9\textwidth]{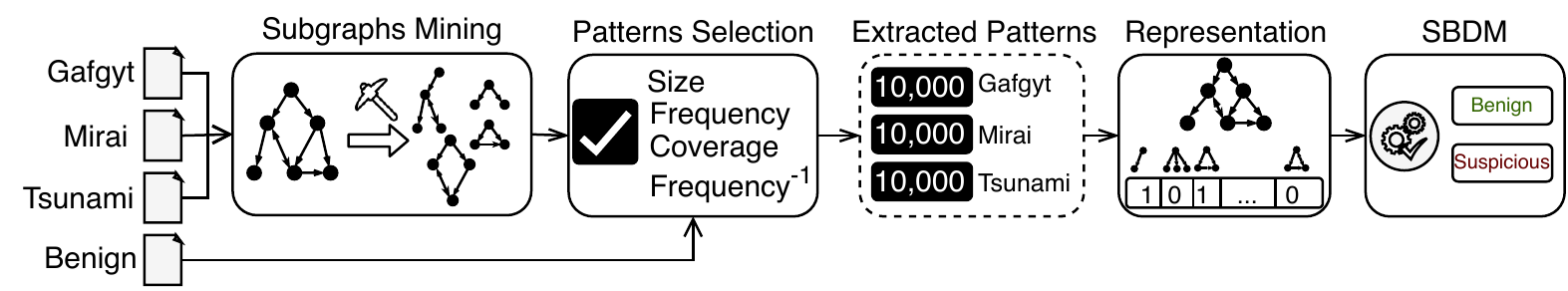}
\caption{Suspicious behavior detection system design. The design consists of four modules, subgraphs mining module to extract frequent subgraphs from three IoT malicious families. Afterward, the subgraphs are ranked by the pattern selection module, where the top 10,000 patterns of each malicious family are selected. Further, 1,000 subgraphs are extracted from the CFG of each sample redirected to the Suspicious Behavior Detector, and represented as a vector of size $30,000$. The vector representation is fed to Suspicious Behavior Detector model to be classified into benign and suspicious.}
\label{fig:SBD_Defense}
\end{figure*}

\BfPara{Suspicious Behavior Detector} Suspicious Behavior Detector is a graph mining-based technique to investigate suspicious malicious patterns within  software samples classified as benign. 
\autoref{fig:SBD_Defense} highlights the design of the Suspicious Behavior Detector, which  consists of four modules, \cib{1} subgraphs mining, \cib{2} pattern selection, \cib{3} data representation, and \cib{4} suspicious behavior detection model. In the following, we describe each module.

\noindent\cib{1} \textbf{Subgraphs Mining:} 
This module extracts common subgraphs within each IoT malware family. Using gSpan, we extracted and collected frequent subgraphs of a size range between 5 to 20 nodes from each malicious family. 
In particular, we used the gSpan algorithm to extract subgraphs from the training samples of each malicious family. This process took more than 160 hours to finish and resulted in over 2,150,170 patterns distributed as 22,953 for Gafgyt, 127,217 for Mirai, and over 2,000,000 for Tsunami families. 
The extracted frequent subgraphs (patterns) for each malicious family are then subjected to further analysis.

\noindent\cib{2} \textbf{Pattern Selection:}
This module ranks the extracted patterns based on four factors: pattern size, frequency, coverage, and inverse frequency. 
For examples, large patterns are assigned higher value since they are distinctive and more likely to be unique to their family. 
Moreover, large patterns can be further decomposed to smaller patterns. 
Further, the number of occurrences of a pattern within a malicious family is considered as an indication of its maliciousness. On the other hand, less frequent patterns are more likely to be function-oriented and solely contribute to the functionality of the code rather than the general behavior of the malware family. 
Therefore, we excluded all patterns that occurred in less than 5\% of the targeted family samples. 
The coverage of the pattern is defined as $\sum_{i=1}^{n} {1}/\text{occurrence}_i-1$, where $n$ is a set of samples in which the pattern occurred, and $occurrence_i$ is the number of patterns contained within the sample $i$.
For example, if a sample contains only one pattern, that pattern will have the highest rank. 
In addition, we compute the number of occurrences for each pattern in the benign training samples. Note that benign samples may have patterns similar to the ones in the malware due to the abstract nature of the CFG and the considered size and functionality of patterns.
To ensure that all patterns hold some behavioral characteristics of the malicious family,  
we excluded all malicious patterns that appeared in more than ten benign samples.
We filtered the patterns and selected the top 10,000 ranked patterns from each family to be its representative patterns. This results in a total of 30,000 malware patterns for the three malware families. We denote this set of patterns as $P$.

\noindent\cib{3} \textbf{Data Representation:}
To investigate an IoT software, we find whether each of the selected 30,000 patterns is a subgraph of the CFG of the software using the VF2 subgraph isomorphism algorithm \cite{CordellaFSV04}. Each sample is represented as a binary vector in the space of the patterns extracted in the previous module, i.e., $v \in \{0,1\}^{|P|}$.
Specifically, we represent each sample by a hot-encoding vector $v$ of size $\text{30,000}$, where $v_i = 1$ if the $i^{th}$ pattern is a subgraph of the sample's CFG, i.e., $v_i = 1 ~\text{if}~~  p_i \subseteq G \ ,\  p_i \in P$. Time-wise, representing a software's CFG as a hot-encoding vector may require several minutes and up to several hours, depending on its size (number of nodes) and structure.

\noindent\cib{4} \textbf{Suspicious Behavior Detection Model:}
Suspicious Behavior Detector model is a CNN-based model trained on the feature representations  extracted from the training dataset shown in \autoref{DSDist}. 
The goal of this module is to investigate suspicious behavior within the sample. 
If the sample is classified as suspicious, further analysis is required by an analyst or dynamic analysis approach.

\BfPara{Experimental Setup}
We trained the Suspicious Behavior Detector model on the feature representation of the training dataset, where all malicious samples are labeled as suspicious. We did not incorporate AEs in the training process, and thus the process may be biased in evaluating the system against samples generated using the same approach (\eg SGEA). Further, we generated AEs using both GEA and SGEA, to force the detection model to misclassify the IoT malware samples as benign, thereby, directing the samples to the Suspicious Behavior Detector component. The generation of the AEs are similar to the process discussed in~\textsection\ref{sec:No_defense}. However, we focus on the malware to benign misclassification only, as such behavior have critical consequences on the system behavior.

\subsection{\fhmc{}: Evaluation and Results}





\BfPara{\fhmc{}: Detection Task}
The \fhmc{} system aims to establish a robust malware classification approach through hierarchical levels of abstractions. The first task of the system is the malware detection, where samples are classified into malicious and benign. This task is conducted using the same model architecture described in \textsection\ref{sec:NDsystemDes} and illustrated in \autoref{fig:CNN_Architecture}. 
The baseline performance of the detection model is shown in \autoref{DetectionSystems} achieving an accuracy of 98.96\%.
We also demonstrated the robustness assessment of the model against AEs generated by GEA and SGEA in ~\autoref{GCNodesMB} and ~\autoref{SGEADetectionM2B}, respectively.
We assume using the same detection model, and evaluate the overall system (\fhmc), by including the performance of the the following components of the system, which are the Suspicious Behavior Detector and the malware classification model.

\begin{table}[t]
    \centering
    \caption{\fhmc{} classifier evaluation on GEA and SGEA AEs classified as malware by the detection model. All SGEA, Tsunami, and large GEA AEs were misclassified to benign, represented as ``--''. Here, all Mirai samples that failed to misclassify the detector where correctly classified.}
    \label{ClassifierAdvBehavior}
    \begin{tabular}{c|c|c|c|c}
        \toprule
        \multicolumn{2}{c|}{Type} & Gafgyt & Mirai & Tsunami \\
        \midrule
        \multirow{3}{*}{GEA} & Small & 65.90\% & 100\% & --  \\ 
        & Median & 50.00\% & 100\% & --  \\ 
        & Large & -- & -- & --  \\ 
        \midrule
        \multicolumn{2}{c|}{SGEA} &  -- & -- & --   \\
        
        \bottomrule
    \end{tabular}
\end{table}

\BfPara{\fhmc{}: Classification Task} 
Malware samples that are detected by the malware detection model, \ie their malicious behavior is successfully detected, are then directed to a classification model to predict the family of the malware, \ie Gafgyt, Mirai, and Tsunami. This model is similar to the classification model in \textsection\ref{sec:NDsystemDes}, however, in this component only malicious samples are classified (no benign class).
To build and evaluate the model for this task, we trained a CNN-based model on malware samples of the three malware families from the training set of our dataset. 
The model achieves an overall accuracy of 97.34\% on malware samples.
To evaluate the model's robustness, we examine the model performance against AEs that fail to fool the detection model, \ie malware AEs that are still detected as malware.
\autoref{ClassifierAdvBehavior} shows the performance of the classification model on the AEs.
Since the malicious AEs generated by SGEA and large GEA are passed with no detection, \ie their misclassification rate is 100\% on the detection task, they were not directed to the classification model and therefore are not included in the results. For the same reason, all samples from the Tsunami family are not included.
However, successfully-detected malicious AEs, \ie belonging to GEA-based samples (small and median), are classified with an accuracy of 100\% for Mirai AEs and 65.90\% and 50.00\% of the Gafgyt AEs generated by small and median GEA approach, respectively.  

\begin{table}[t]
    \centering
    \caption{\fhmc{} Suspicious Behavior Detector evaluation on benign and adversarial samples. DO refers to Data origin, and AR is the accuracy rate. Here, 93.00\% of the benign samples are correctly classified. Additionally, 100\% of the large GEA AEs are detected as suspicious, with an overall performance of 90.79\%.}
    \label{SBDDetector}
    \begin{tabular}{c|c|c|c|c|c|c}
        \toprule
        \multirow{2}{*}{DO} & \multirow{2}{*}{Benign} & \multicolumn{3}{c|}{GEA} & \multirow{2}{*}{SGEA} & \multirow{2}{*}{Overall} \\
        & & Small & Median & Large & & \\
        \midrule
        AR & 93.00\% & 62.79\% & 98.78\% & 100\% & 79.63\% & 90.79\%\\
        \bottomrule
    \end{tabular}
\end{table}

\BfPara{\fhmc{}: Suspicious Behavior Detection Task}
This component aims to further investigate the benign-classified samples based on patterns extracted from their structural components. The task of Suspicious Behavior Detector, \ie CNN-based classifier built on features that represent the existence of malicious patterns within the input sample, is to determine whether a given sample is signaling a suspicion of malicious behavior, and therefore it is operating as an AEs detection technique.
We evaluate the Suspicious Behavior Detector 
using the original benign samples and malicious AEs. 
As shown in~\autoref{SBDDetector} the detector achieves an accuracy of 93.00\% for classifying benign samples and an accuracy of 62.79\%, 98.78\%, and 100\% for classifying GEA-based AEs with small, median, and large perturbation, respectively. The detector achieves a detection accuracy of 79.63\% for AEs generated using SGEA approach. Introducing large perturbations in the samples increases the changes of introducing some subgraph patterns that are signatures of the malicious samples. This explains the high accuracy achieved by the detector for large perturbations.
This results show that using \fhmc{} enabled a systematic methods of coping with adversarial manipulation to malware. When a suspicious behavior is detected for a given sample, other methods can be adopted to analyze the sample in order to provide robust evaluation of malicious activities.

\section{Discussion}\label{sec:discussion}
\BfPara{\fhmc{}: Cost of Security}
The security of machine learning algorithms is important for many applications since it has been shown that such algorithms are vulnerable to adversarial attacks (\ie AEs). In malware classification field, AEs pose critical security implications as emerging studies have shown that AEs can fool the machine learning-based malware detection system \cite{AbusnainaKAPAM,AbusnainaAASNM19,SuciuCJ19}. However, 
limited studies have investigated potential defenses. 
In this work, we show that launching adversarial attacks against malware detection system can lead to a misclassification rate of as high as 100\%. 
To cope with such adversarial settings and capabilities, we introduced \fhmc{} that operates on multiple level of behavioral analysis of the software to insure its security.
Since AEs are derived from a combination of benign and malware components, detecting them is a challenging task and often comes at the cost of misclassifying a portion of benign samples as malicious, hence producing false alarms.
For example, in our experiments, we show that successfully-detecting 
88.52\% of AEs is accompanied by 7.00\% false alarms resulting from detecting suspicious behavior in the benign samples.

\BfPara{\fhmc{} Robustness}
The suspicious behavior detector in \fhmc{} consists of a CNN-based model trained on the vector representations of the extracted patterns. In the context of adversarial attacks, it should be robust against perturbation to the original samples. Otherwise, it will suffer from the effects of adversarial examples. We took this into account in the design process. Ideally, the detector represents the benign samples into a vector representation of zeros. In our dataset, $\approx$ 85\% of the benign samples are represented as a vector of all zeros, indicating that none of the malicious patterns captured within them. Therefore, to generate successful AEs against \fhmc, adversaries have to conceal malicious patterns by modifying the functionality of the malicious sample, therefore, contrasting the practicality and functionality requirements of the generated AE.

\BfPara{Potential Investigation Technique: Dynamic Analysis}
For accurate and fail-proof malware detection, every sample should be analyzed dynamically and filtered based on its behavior. However, dynamic analysis has its own downsides: \cib{1} It requires setting up of a sandboxed environment such that the execution of the malware does not impact the underlining host system. \cib{2} It is costly in terms of time and memory. 
These make the dynamic analysis techniques difficult to scale. Our proposed technique puts forward a static analysis based fine-grained hierarchical approach towards malware detection. The samples classified as benign in the first phase are sent to the Suspicious Behavior Detector for investigation using the deeper CFG based features (subgraphs). The samples that are detected as suspicious in the second phase are finally considered as potential for dynamic analysis. Statistically, only 2.42\% (42 out of 1,733) of the normal samples, \ie non-adversarial, were identified as the ones to be directed for dynamic analysis, thereby reducing the load on the dynamic analysis technique, hence overcoming its potential difficulties.



\BfPara{Binary Obfuscation} Malware authors often use different packing techniques, \eg Ultimate Packer for Executables (UPX), to obfuscate different parts of the malware code base, such as functions and strings. In obfuscated functions, the CFG would differ from the actual unpacked malware. Thereby, the detector should be aware of obfuscation and can accurately classify the obfuscated software and examining the behavior of packed and unpacked software.

\section{Related Work}\label{sec:Related_Work}
\BfPara{Malware Analysis}
While a lot of effort has been put towards malware analysis and detection in general, IoT malware analysis still lacks the effort. Among the IoT malware studies, efforts towards the analysis and detection of malicious software are limited, particularly, from the lens of CFG. ManXu~\etal\cite{XuRQC18} proposed a CNN-based malware detection system for the Android application from the semantic representation of the graph (\ie control and data flow graphs representations). In addition, Yang~\etal\cite{YangXGYP14} identified and detected Android malicious behaviors throughout generating two-level behavioral representations built from the CFG graph and call graphs of the program. Allix~\etal\cite{AllixBJKST16} designed multiple machine learning classifiers to detect Android malware using different textual representations extracted from the applications' CFGs. Further, Alasmary~\etal\cite{AlasmaryA0CNM18} conducted an in-depth CFG-based comparative study for the Android and IoT malware. 
Similarly, Pa~\etal\cite{PaSYMKR16} established the first IoT honeypot and sandbox system, called IoTPOT, that run over eight CPU architectures to capture the IoT attacks running over Telnet protocol. Similarly, Caselden~\etal\cite{CaseldenBPMS13} built an algorithm that generates an attack from the representation of the hybrid information and CFG applied to the program binaries. Alam~\etal\cite{AlamHTS15} proposed a metamorphic malware analysis and detection system that uses two different techniques that match the CFGs of small malware and then address the change in the opcodes frequencies. Moreover, Tamersoy~\etal\cite{TamersoyRC14} proposed a malware detection algorithm that identifies the executable files of the malware and then computes the similarities between them to partial dataset files from the Norton Community Watch. Then, they construct graphs based on the measurement of inter-relationship between these files. In addition, Wuchner~\etal\cite{WuchnerOP15} proposed a graph-based detection system that uses a quantitative data flow graphs generated from the system calls, and use the graph node properties, i.e., centrality metric, as a feature vector for the classification between malicious and benign programs. Moreover, they extended the work by using a compression-based mining technique applied to the quantitative data flow graphs for malware detection~\cite{WuchnerCOP19}. Moreover, Cen~\etal\cite{CenGSL15} used Android API calls as features extracted from the decompiled source code of the software, and proposed a probabilistic logistic regression-based model for malware detection.

\BfPara{Adversarial Machine Learning}
Machine/deep learning networks are widely used in security-related tasks, including malware detection~\cite{antonakakis2012throw, MohaisenAM15, MohaisenA13, AlasmaryKAPCAANM19}. However, it has been shown that deep learning-based models are vulnerable against adversarial attacks~\cite{miyatoMKNI15}. Given that, it should be noted that such a behavior can be a critical issue in malware detection systems, where misclassifying malware as benign may result in disastrous consequences~\cite{AbusnainaAASNM19,AbusnainaMYM19}. Various adversarial machine learning attack methods in the context of image classification have been introduced to generate AEs. For example, Goodfellow~\etal~\cite{GoodfellowSS15} introduced FGSM, a family of fast method attacks to generate AEs that forces the model to misclassification. In addition, Carlini~\etal~\cite{Carlini017} proposed three L-norm-based adversarial attacks, known as C\&W adversarial attacks, to investigate the robustness of neural networks and existing adversarial defenses. Similarly, Moosavi~\etal~\cite{Moosavi-Dezfooli16} proposed DeepFool, an $L_2$ distance-based adversarial iterative method to generate AEs with minimal perturbation. 
Further, a critical application of the AEs is malware detection. Recent studies investigated generating AEs in the context of malware detection~\cite{SuciuCJ19}. For instance, Grosse~\etal~\cite{GrossePMBM17} implemented an augmented adversarial crafting algorithm to generate AEs, misleading a CNN-based classifier to misclassify 63\% of the malware samples to benign. The detection of the AEs is challenging~\cite{Carlini017_2}. While, to the best of our knowledge, no one has investigated the detection of AEs in the context of malware detection, multiple studies attempt to detect them in the context of image classification~\cite{Xu0Q18,LiL17,MetzenGFB17}, achieving detection accuracy of 20\% to 80\%.
In this study, we implemented \fhmc{}, a graph mining-based fine-grained hierarchical learning approach for suspicious behavior detection, achieving an accuracy of 88.52\% in detecting CFG-based AEs.

\section{Conclusion}\label{sec:conclusion}

This work investigates the robustness of graph-based deep learning models against adversarial machine learning attacks. To set out, first, an in-depth analysis of malware binaries is conducted through constructing abstract structures using CFG, which are analyzed from multiple aspects, such as the number of nodes and edges, as well as graph algorithmic constructs, such as average shortest path, betweenness, closeness, density, etc. Then, two IoT malware classification approaches are introduced, \ssmc{} and \fhmc{}. Then, we generate AEs using two state-of-the-art adversarial attacks, GEA and SGEA approaches. We observed that both GEA and SGEA are capable to misclassify 100\% of malware samples as benign in \ssmc. However, \fhmc{} was able to detect 88.52\% of the AEs, improving the overall robustness in compare to \ssmc{}.

\balance
\bibliographystyle{acm}
\bibliography{ref}

\end{document}